\documentclass[twocolumn,trackchanges]{aastex701}

\usepackage{newtxtext}          
\usepackage{newtxmath}
\usepackage{amsmath}	        
\usepackage{textcomp,gensymb}	
\usepackage{xcolor}             

\newcommand\ms{\ensuremath{\text{m}\,\text{s}^{-1}}}
\newcommand\kms{\ensuremath{\text{km}\,\text{s}^{-1}}}
\newcommand\masyr{\ensuremath{\text{mas}\,\text{yr}^{-1}}}

\newcommand{\thisstar}{HD~38230}
\newcommand{\thisstarA}{HD~38230~A}
\newcommand{\thisstarB}{HD~38230~B}

\definecolor{my_color}{HTML}{CF0000}

\accepted{to AJ}

\begin{document}

\title{A New Sirius-like System at Only 21 Parsecs: An Elusive White Dwarf Companion to the Nearby K-dwarf HD 38230}

\correspondingauthor{Alexander Venner}

\author[0000-0002-8400-1646]{Alexander Venner}
\affiliation{Max Planck Institute for Astronomy, 69117 Heidelberg, Germany}
\email[show]{alvenner@mpia.de}

\author[0000-0003-3904-7378]{Logan A. Pearce}
\altaffiliation{These authors contributed equally to this work.}
\affiliation{Department of Astronomy, University of Michigan, Ann Arbor, MI 48109, USA}
\email{lapearce@umich.edu}

\author[0000-0003-0115-547X]{Qier An}
\altaffiliation{These authors contributed equally to this work.}
\affiliation{Department of Physics, University of California Santa Barbara, Santa Barbara, CA 93106, USA}
\affiliation{Department of Physics and Astronomy, Johns Hopkins University, Baltimore, MD 21218, USA}
\email{qan4@jh.edu}

\author[0000-0002-2428-9932]{Cl\'emence Fontanive}
\affiliation{SUPA, Institute for Astronomy, University of Edinburgh, Edinburgh, EH9 3HJ, UK}
\email{clemence.fontanive@roe.ac.uk}

\author[0000-0003-0593-1560]{Elisabeth C. Matthews}
\affiliation{Max Planck Institute for Astronomy, 69117 Heidelberg, Germany}
\email{matthews@mpia.de}

\author[0000-0003-4557-414X]{Kyle Franson}
\altaffiliation{NHFP Sagan Fellow}
\affiliation{Department of Astronomy and Astrophysics, The University of California Santa Cruz, Santa Cruz, CA 95064, USA}
\email{kfranson@utexas.edu>}

\author[0000-0003-0800-0593]{Justin R. Crepp}
\affiliation{Department of Physics and Astronomy, University of Notre Dame, Notre Dame, IN 46556, USA}
\email{jcrepp@nd.edu}

\author[0000-0001-9608-6395]{Hai Fu}
\affiliation{Department of Physics \& Astronomy, The University of Iowa, Iowa City, IA 52242, USA}
\email{hai-fu@uiowa.edu}

\author[0000-0003-2630-8073]{Timothy D. Brandt}
\affiliation{Space Telescope Science Institute, Baltimore, MD 21218, USA}
\affiliation{Department of Physics, University of California Santa Barbara, Santa Barbara, CA 93106, USA}
\email{tbrandt@stsci.edu}

\begin{abstract}

White dwarfs are among the rarest and faintest stars found in the solar neighbourhood and are therefore persistently challenging to detect. Whereas the \textit{Gaia} mission has vastly expanded our knowledge of isolated white dwarfs, limited contrast sensitivity means that white dwarfs companions to much brighter solar-type stars are easily missed, leaving a subset of white dwarfs in ``Sirius-like systems" that can only be discovered via high-contrast imaging. Here we report the discovery of a white dwarf companion to the nearby K0~V star HD~38230 (HIP 27207). The companion, HD~38230~B, was originally detected with Keck/NIRC2 observations taken as part of the TRENDS high-contrast imaging survey, was detected by \textit{Gaia} without an astrometric solution, and was independently observed using Lick/ShaneAO and Palomar/PHARO in the literature. Combining these multiple imaging detections spanning 6~years, we demonstrate that HD~38230~B is a gravitationally bound companion and has photometry consistent only with a white dwarf. Our analysis of the binary orbit, combining relative astrometry with more than 25~years of precise radial velocity observations and \textit{Hipparcos-Gaia} astrometry, results in strong constraints on the $P = 1390^{+310}_{-200}$~year orbit despite the short observational span, and provides a precise dynamical mass of $M_B=0.71^{+0.06}_{-0.05}~M_\odot$ for the white dwarf. At 21~pc, this is the 11$^\text{th}$-nearest Sirius-like system known to date, highlighting residual incompleteness in the local white dwarf census.


\end{abstract}


\keywords{white dwarf stars --- binary stars --- direct imaging --- radial velocity --- astrometry --- orbit determination}


\section{Introduction} \label{sec:intro}

White dwarfs (WDs) are the dim remnants of stars left behind after the ravages of stellar evolution. They are typically much fainter than their main-sequence progenitors, which leads to considerable challenges in their detection, especially when they are found in close binaries. For this purpose, \citet{Holberg2013} defined the term ``Sirius-like system" \citep[SLS, originally coined by][]{Barstow2001} as encompassing binary and multiple-star systems incorporating a white dwarf with a non-degenerate component with spectral type earlier than M. The observational rationale for this is that while a white dwarf and an M-dwarf may have comparable luminosity in visible wavelengths, companion stars with earlier spectral types will quickly outshine the white dwarf component, making them much more challenging to detect. Nonetheless, the three earliest white dwarfs to be discovered, Sirius~B, Procyon~B, and 40~Eridani~B, are all found in Sirius-like systems, demonstrating the significant role of these systems in our understanding of white dwarfs (see \citealt{Holberg2009} for a review).

Sirius-like systems are observationally valuable for several reasons. For example, most of the small sample of model-independent dynamical masses for white dwarfs have been derived from nearby Sirius-like systems \citep[e.g.][]{Bowler2021, Zhang2023}. Furthermore, non-degenerate companions to the white dwarfs provide information on metallicity, otherwise lost for the white dwarf, allowing for further study such as investigation of metallicity dependence of white dwarf metal pollution \citep{Jenkins2024}. Sirius-like systems also offer a means to constrain the total ages of field white dwarfs via their non-degenerate companions, allowing for estimation of the progenitor masses which, in turn, constrains the white dwarf initial-final mass relation \citep{Catalan2008, Zhao2012, Barrientos2021}; this can also be inverted to use white dwarfs to estimate ages of their main sequence companions \citep[e.g.][]{Fouesneau2019}. Furthermore, for systems where the ages of the white dwarf and non-white dwarf components can both be constrained, Sirius-like systems can also be used to test details of white dwarf cooling models \citep{Venner2023, Barrientos2024}.

Despite their value for understanding white dwarfs, Sirius-like systems are unfortunately quite challenging to detect. For a Sun-like primary star the flux contrast of a white dwarf companion in visible wavelengths is typically in the region of $10^{-4}$ ($\Delta m~\approx$ 10~mag) -- and usually even greater in the infrared -- which was historically an extremely challenging level of flux sensitivity to achieve. In their review, \citet{Holberg2013} were aware of only 98 Sirius-like systems, a sample which appeared to be substantially incomplete beyond 20~pc.

In the past decade, the \textit{Gaia} mission \citep{Gaia} has revolutionised the field of white dwarf research, primarily through the vast increase in the number of known white dwarfs based on the precise all-sky astrometry and photometry \citep{Tremblay2024}. This has, in turn, led to a major increase in the number of white dwarfs in resolved binaries, unlocking new understanding of their physics \citep[e.g.][]{ElBadry2018}. However, primarily due to the large flux contrasts involved, white dwarf companions to solar-type stars are difficult to detect with \textit{Gaia} for projected separations below $\lesssim$5\arcsec{} \citep[see e.g.][]{Venner2025}. As a result, high-contrast imaging remains a vital tool for assembling a complete sample of Sirius-like systems.

As white dwarfs tend to have comparable masses to their main-sequence companions, their gravitational influence within binaries is often more obvious than their flux contribution. Many of the white dwarfs in nearby Sirius-like systems have been discovered via high-contrast imaging, often specifically motivated by radial velocity (RV) trends \citep{Crepp2013b, Crepp2018, Zurlo2013, Rodigas2016, Hirsch2019, Bowler2021} or astrometric accelerations \citep{Bonavita2020} generated by the WD companion. These observations are usually conducted in the infrared, since high-contrast imaging instruments are typically optimised for detection of cool substellar companions; however, as white dwarfs are often hotter than their companion stars, there is an advantage to attempting to resolve Sirius-like systems at shorter wavelengths. A demonstration of the potential efficacy of this approach is seen in \citet{Pearce2025}, who used visible-wavelength high-contrast imaging on a pre-selected sample of stars with UV excesses suggesting the presence of white dwarf companions, and successfully detected five companions to 18 targets.

In this work, we report the discovery of a new white dwarf companion to the nearby ($D$ = 20.880$\pm$0.009~pc) K0~V star \thisstar{}. The white dwarf, \thisstarB{}, is detected both by its gravitational influence on \thisstarA{} (evident in RVs and \textit{Hipparcos-Gaia} astrometry), and spatially resolved at a projected separation of $\rho\approx4.7$\arcsec{} ($\approx$100~AU). The combination of direct and indirect observational methods allows us to constrain the orbit of the binary, despite its long orbital period ($>$1000~yr), and provides a precise dynamical mass for the white dwarf. This discovery demonstrates that the census of white dwarfs in Sirius-like systems remains incomplete even at relatively close distances to the Sun.

\section{Data} \label{sec:data}

\subsection{Imaging} \label{subsec:imaging}

\subsubsection{Keck/NIRC2}

\begin{figure}[t!]
\centering
\includegraphics[width=0.48\textwidth]{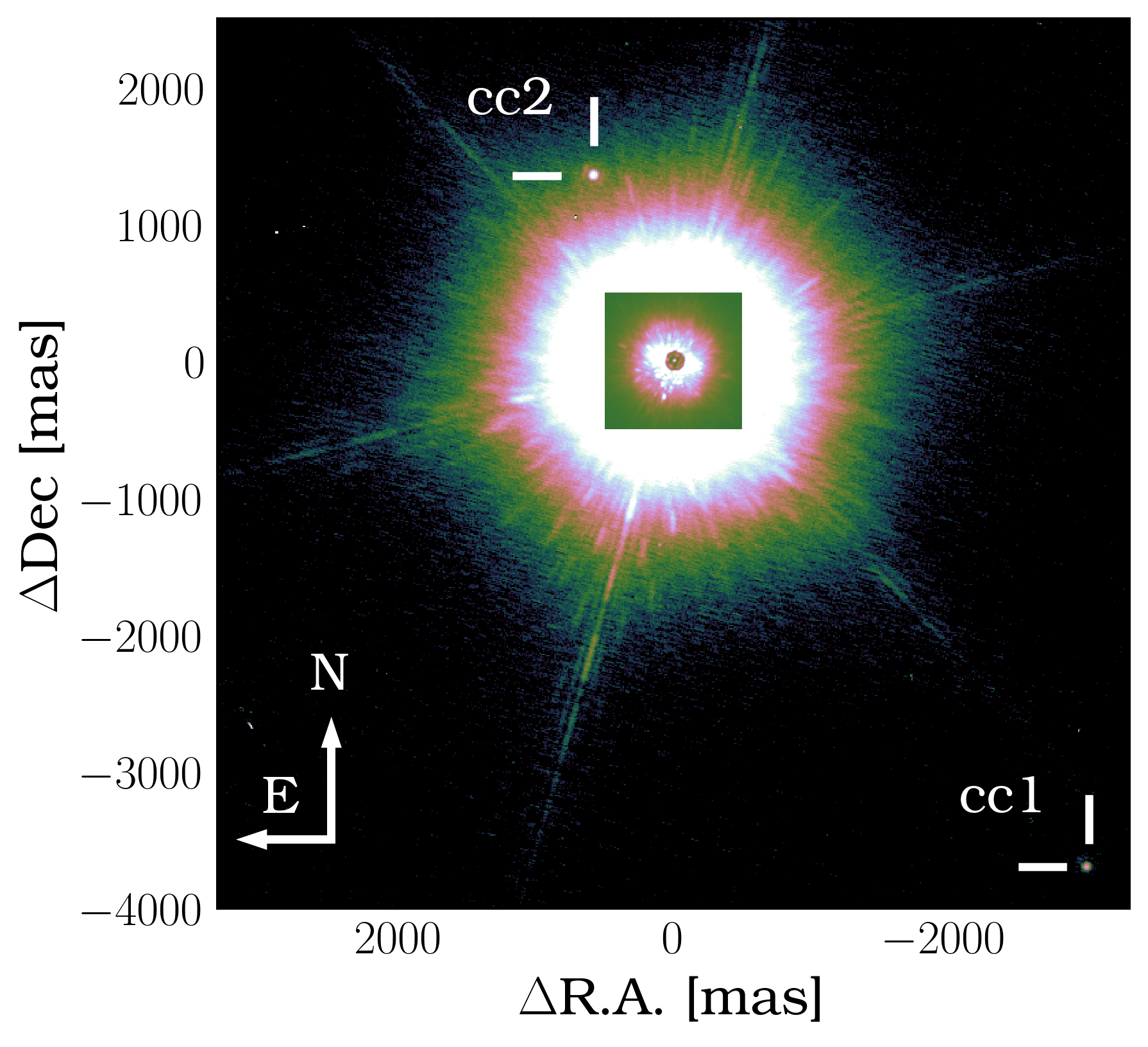}
\caption{\small{Keck/NIRC2 adaptive optics image of HD 38230 from 2011 Feb 22. The central star is shown in a linear stretch (inset) and is behind the 300 milliarcsecond (mas) Lyot coronagraph; the background is shown with a square root stretch to emphasize the faint companions. In this image the white dwarf, labelled CC1, is located to the southwest at separation and position angle [4.75", 219.0$\degree$], while a second star in the field, CC2, is located to the northeast at [1.47", 23.7$\degree$]; both are marked with white crosshairs. X and Y axes reflect the spatial offset from HD 38230 in mas.}}
\label{fig:NIRC2image}
\end{figure}

\thisstar{} was observed with the near-infrared imaging camera NIRC2 \citep{Wizinowich2000} on the Keck-II as part of the TaRgetting bENchmark-objects with Doppler Spectroscopy (TRENDS) imaging survey, which targeted stars displaying long-term Keck/HIRES RV accelerations (see Section~\ref{subsec:RV}) using high-contrast imaging observations to search for wide-separation companions \citep{Crepp2012b, Crepp2013a, Crepp2014, Crepp2016, Montet2014, Gonzales2020}. A total of three NIRC2 epochs were obtained in natural guide star AO mode on 2011 Feb 22 and 2012 Feb 02 in the $K^\prime$ band and on 2015 Sep 24 in $K_s$.
All images were obtained with the full-frame (1024$\times$1024 pixels) narrow camera in vertical angle mode (which allowed the sky to rotate in the image plane during observation) with the star behind the 300 milliarcsecond (mas) Lyot coronagraph mask. We accessed the science images and relevant calibration files from the Keck Observatory Archive\footnote{\url{https://koa.ipac.caltech.edu/}} via the PyKOA python client\footnote{\url{https://koa.ipac.caltech.edu/UserGuide/PyKOA/TAPClients.html}}.

Figure \ref{fig:NIRC2image} shows a reduced and derotated image from the 2011 epoch (reduction described in Section~\ref{subsec:NIRC2_analysis}). Two companion candidates (CC) can be seen in all three NIRC2 observations, marked with white crosshairs here. The more widely-separated, labelled CC1, lies to the south-west of \thisstar{} at separation and position angle [4.75", 219.0$\degree$], while the second, labelled CC2, lies to the north-east at [1.47", 23.7$\degree$]. Further analysis of the companion candidates is carried out in Section~\ref{subsec:NIRC2_analysis}.

\subsubsection{Literature Imaging} \label{subsec:lit_imaging}

Three epochs of high contrast imaging of \thisstar{} using Lick/ShaneAO and Palomar/PHARO dated to 2016 Jan 21, 2016 Oct 20, and 2017 Jan 17 were published in \citet{Hirsch2021}. Both of the aforementioned companion candidates can be identified in their observations, extending the total time baseline of observations forwards by about one additional year. The interpretation of results for \thisstar{} are not discussed in detail in \citet{Hirsch2021}, however both sources were ultimately flagged as background stars.

We utilise the relative astrometry published in \citet{Hirsch2021} for this work. However, a direct comparison of their values and the NIRC2 astrometry derived in this work show larger-than-expected differences. In particular, the ShaneAO position angles appear to differ substantially. Regarding the position angle, \citet{Hirsch2021} report an instrument field rotation of $1.87\pm0.1\degree$; we find that adding this offset to the published position angles brings them into close agreement, so we apply this rotation to the ShaneAO position angles for this work. However, as the residuals of the archival data remain high, we ultimately resort to inflating the uncertainties on the relative astrometry for the final orbital model (Section~\ref{subsec:orbit_model}). \citet{Hirsch2021} also report relative photometry for both companion candidates. The Palomar/PHARO observations provide $\Delta K_s=9.80\pm0.20$~mag for CC1 and $\Delta K_s=9.10\pm0.13$~mag for CC2 relative to the primary star.

\subsubsection[Gaia]{\textit{Gaia}} \label{subsec:Gaia}


\thisstarA{} is included in \textit{Gaia}~DR3 with the source ID 3456337827237371264. The \textit{Gaia}~DR3 astrometric solution has a parallax of 47.855$\pm$0.020~mas, corresponding to a distance of 20.880$\pm$0.009~pc \citep{BailerJones2021}, and a high proper motion of $\mu$ = [+486.410$\pm$0.022, --510.139$\pm$0.013]~\masyr{}.
Within an 8" radius of \thisstarA{}, only two additional sources are detected in \textit{Gaia}~DR3; one at a projected separation and position angle of [4.72", 218$\degree$] with source ID 3456337827238475008, and one at [4.22", 335$\degree$] with source ID 3456337827238475008. By comparison with the most temporally proximate NIRC2 and ShaneAO observations, these can confidently be identified with CC1 and CC2 respectively.

In \textit{Gaia}~DR3, CC2 has a $G$-band magnitude of 16.68 and has a 5-parameter astrometric solution with parallax $\varpi$ = 3.14$\pm$0.22~mas and proper motion $\mu_\alpha,\mu_\delta$ = [+3.21$\pm$0.28, -0.71$\pm$0.15]~\masyr{}. Though the Renormalised Unit Weight Error (RUWE) parameter for the astrometric solution has a value of 2.2, suggesting a comparatively poor fit to the \textit{Gaia} astrometry, the parallax and proper motion strongly suggest that CC2 is a background star unrelated to \thisstarA{}. On the other hand, in \textit{Gaia}~DR3 CC1 only has a two-parameter solution, with no constraint on its parallax and proper motion at all \citep{Lindegren2021}, despite the fact it is brighter than CC2 ($G$ = 15.79). There does not appear to be an internal reason why a \textit{Gaia} 5-parameter solution may have failed, but this is presumably related to its proximity to the much brighter primary star. As a result, the nature of CC1 cannot be determined from \textit{Gaia}~DR3 alone.

Both CC1 and CC2 have colour BP/RP photometry in \textit{Gaia}~DR3, which is valuable for interpreting their physical nature, with nominal BP-RP values of 0.72 and 1.51 respectively. Yet these show strong evidence for poor quality, with the \texttt{phot\_bp\_rp\_excess\_factor} parameter having high values of 3.90 and 4.57 for CC1 and CC2. Following \citet[][equation 6]{Riello2021}, this results in respective normalised flux excess factors $C^*$ of 2.70 and 3.29, which both greatly exceed the expected value of 0 for well-behaved sources. The most likely explanation for the poor quality of the \textit{Gaia} colour photometry is flux contamination from the $\sim$10 mag brighter primary star. Thus we conclude that the \textit{Gaia} BP/RP colour photometry for CC1 and CC2 is likely to be unreliable.

\subsection{Radial Velocities} \label{subsec:RV}

\thisstarA{} has long-term high-precision RV observations with the Keck/HIRES precision RV spectrograph. For this work we use the most recent release of HIRES RV data in the literature from \citet{Teklu2025},  which builds upon the work from \citet{Butler2017} which included the first public release of HIRES RVs for this star. The HIRES dataset published in \citet{Teklu2025} includes 114 epochs spanning 9410~days (25.8~yr) between 1996~December~02 and 2022~November~12. HIRES underwent an instrument upgrade in 2004, which induced a $\sim$1.5~\ms{} RV offset; this has been subtracted in the datasets published in \citet{Teklu2025}, along with low-level changes in the nightly RV zero-point. We therefore treat the HIRES RVs as a singular dataset, without a free offset between the pre- and post-upgrade RVs.

In \citet{Butler2017}, \thisstarA{} was identified to display a highly significant +3.800$\pm$0.073~m~s$^{-1}$~yr$^{-1}$ RV acceleration. The same acceleration underlies its selection for NIRC2 observations in the TRENDS imaging survey \citep{Crepp2012b}. Over the course of HIRES RV dataset used in this work, this manifests as a linear drift with an observed $\Delta$RV of $\sim$100~\ms{}. The combination of Keck/HIRES RV accelerations and the NIRC2 imaging detections in such cases offers strong potential to measure precise dynamical masses \citep[see e.g.][]{Crepp2012a}.

\subsection{Absolute Astrometry} \label{subsec:astrometry}

We adopt absolute astrometry of \thisstarA{} (HIP~27207) from the \textit{Hipparcos--Gaia} Catalog of Accelerations \citep[HGCA;][]{Brandt2018, Brandt2021}. The HGCA cross-calibrates the \textit{Hipparcos} and \textit{Gaia} astrometric solutions onto a common reference frame and applies a careful error budget to achieve statistically well-behaved uncertainties for each proper-motion measurement. The catalog provides three distinct proper motions: one from \textit{Hipparcos}, one from \textit{Gaia}, and a long-baseline mean proper motion derived from the difference in positions between the two missions divided by the $\sim$25~yr time baseline. Differences among these three proper motions indicate acceleration in an inertial reference frame, signalling the gravitational influence of a companion.
 
\thisstarA{} shows a modest, yet highly significant proper motion anomaly in the HGCA. The \textit{Gaia} proper motion differs from the long-term \textit{Hipparcos--Gaia} proper motion by $\Delta\mu=[+0.240\pm0.042,+0.260\pm0.025]$~\masyr{}, resulting in a high $\chi^2$ of 154.07 for a constant proper motion model. This drift in the proper motion is equivalent to a net change in tangential velocity of $\Delta v_\text{tan}=[+24\pm4,+26\pm2]$~\ms{}. This strongly suggests the presence of a hitherto unknown companion to \thisstar{}. On the other hand, the \textit{Gaia} astrometry itself suggests that this companion likely has a long orbital period; \thisstarA{} has a RUWE of 0.90 in \textit{Gaia}~DR3, indicating that a 5-parameter single-star astrometric solution provides a good model for the underlying \textit{Gaia} astrometry. This can be understood given a companion with an orbital period much longer than the $\sim$3~yr \textit{Gaia}~DR3 observational baseline, resulting in nearly linear reflex motion over the course of observations that can be absorbed into the linear proper motion terms rather than degrading the astrometric residuals \citep{An2025}.

\section{Method} \label{sec:method}

\subsection{Properties of HD 38230 A}

\begin{deluxetable*}{lrr}[ht!]
\label{tab:star}
\centering
\tablecaption{Properties of \thisstarA{}.}
\tablehead{\colhead{Parameter} & \colhead{Value} & \colhead{Reference}}
\startdata
~~Right Ascension $\alpha_{\text{J2000}}$ (*) \dotfill & 05:46:01.89 & \citet{GaiaDR3} \\
~~Declination $\delta_{\text{J2000}}$ (*) \dotfill & +37:17:04.74 & \citet{GaiaDR3} \\
~~$V$ (mag) \dotfill & $7.341\pm0.012$ & \citet{Tycho2} \\
~~Parallax $\varpi$ (mas) \dotfill & $47.855\pm0.020$ & \citet{GaiaDR3} \\
~~Distance $D$ (pc) \dotfill & $20.880\pm0.009$ & \citet{BailerJones2021} \\
~~R.A. proper motion $\mu_\alpha$ (\masyr{}) \dotfill & $+486.410\pm0.022$ & \citet{GaiaDR3} \\
~~Declination proper motion $\mu_\delta$ (\masyr{}) \dotfill & $-510.139\pm0.013$ & \citet{GaiaDR3} \\
~~Radial velocity (km~s$^{-1}$) \dotfill & $-29.072\pm0.001$ & \citet{Soubiran2018} \\
~~$U$ (km~s$^{-1}$) \dotfill & $+21.50\pm0.10$ & This work \\
~~$V$ (km~s$^{-1}$) \dotfill & $-71.35\pm0.03$ & This work \\
~~$W$ (km~s$^{-1}$) \dotfill & $+12.91\pm0.01$ & This work \\
\hline
~~$B_\text{T}$ (mag) \dotfill & $8.405\pm0.018$ & \citet{Tycho2} \\
~~$V_\text{T}$ (mag) \dotfill & $7.429\pm0.012$ & \citet{Tycho2} \\
~~$G$ (mag) \dotfill & $7.121\pm0.02$ & \citet{GaiaDR3} \\
~~$G_{\text{BP}}$ (mag) \dotfill & $7.539\pm0.02$ & \citet{GaiaDR3} \\
~~$G_{\text{RP}}$ (mag) \dotfill & $6.531\pm0.02$ & \citet{GaiaDR3} \\
~~$J$ (mag) \dotfill & $5.838\pm0.024$ & \citet{2MASS} \\
~~$H$ (mag) \dotfill & $5.460\pm0.027$ & \citet{2MASS} \\
~~$K_s$ (mag) \dotfill & $5.353\pm0.020$ & \citet{2MASS} \\
~~$W1$ (mag) \dotfill & $5.355\pm0.182$ & \citet{AllWISE} \\
~~$W2$ (mag) \dotfill & $5.186\pm0.058$ & \citet{AllWISE} \\
~~$W3$ (mag) \dotfill & $5.380\pm0.016$ & \citet{AllWISE} \\
~~$W4$ (mag) \dotfill & $5.318\pm0.032$ & \citet{AllWISE} \\
\hline
~~Spectral type \dotfill & K0 V & \citet{Gray2003} \\
~~[Fe/H] (dex) \dotfill & $-0.07\pm0.03$ & \citet{Rice2020} \\
~~[Mg/H] (dex) \dotfill & $+0.03\pm0.04$ & \citet{Rice2020} \\
~~$T_{\text{eff}}$ (K) \dotfill & $5170\pm56$ & \citet{Rice2020} \\
~~$M_*$ ($M_\odot$) \dotfill & $0.82\pm0.02$ & This work \\
~~$R_*$ ($R_\odot$) \dotfill & $0.81\pm0.03$ & This work \\
~~log~$g$ ($\log$ cm~s$^{-2}$) \dotfill & $4.54\pm0.03$ & This work \\
~~$L_*$ ($L_\odot$) \dotfill & $0.42\pm0.03$ & This work \\
~~Age (Gyr) \dotfill & $\approx$6 -- 10 & This work \\
\enddata
\tablecomments{(*) Propagated backwards to Equinox J2000 from the \textit{Gaia}~DR3 solution (i.e. assuming the \textit{Gaia}~DR3 proper motion).}
\end{deluxetable*}

\thisstarA{} is a $V=7.3$ star \citep{Tycho2} with spectral type K0~V \citep{Cowley1967, Gray2003} with a distance of $20.880\pm0.009$~pc \citep{BailerJones2021}. \thisstarA{} is therefore among the nearer solar-type stars to the Sun. We report the \textit{Gaia}-based basic properties of \thisstarA{} in Table~\ref{tab:star}.

Owing to its brightness, spectral classification, and nearby location, \thisstarA{} has a strong presence in the astronomical literature. In particular, largely due to its inclusion in the Keck/HIRES RV survey, the spectroscopic properties of the star are well-studied \citep[e.g.][]{Valenti2005, Takeda2007}. In this work we adopt spectroscopic parameters from \citet{Rice2020}, who applied a machine learning model to stars with Keck/HIRES spectra to determine their properties.

As \thisstarA{} is a K-dwarf, its main sequence lifetime is relatively long in comparison to the age of the universe, meaning that stellar models are not strongly sensitive to its age. However, we can derive some constraints on the age of the system from its galactic kinematics. We calculate the space velocities of \thisstarA{} following \citet{Johnson1987}, using the astrometric solution from \textit{Gaia}~DR3 and the absolute radial velocity from \citet{Soubiran2018} as inputs. As these values are instantaneous they will be affected by orbital motion, but given the long orbital period of the binary we expect that this should not significantly affect our inferences.

In this way, we find space velocities of $(U, V, W)=(+21.50\pm0.10, -71.35\pm0.03, +12.91\pm0.01)$~\kms{}, where $U$ is defined as positive in the direction of the galactic centre, $V$ is positive towards the galactic rotation, and $W$ is positive in the direction of the north galactic pole. Adopting the solar space velocities from \citet{Schonrich2010}, we find absolute motions relative to the local standard of rest $(U_{\text{LSR}}, V_{\text{LSR}}, W_{\text{LSR}})=(+32.6\pm0.8, -59.1\pm0.5, +20.2\pm0.4)$~\kms{}. Following \citet{Bensby2003}, this places the star in the ambiguous region between the thin disk and thick disk. However, the total velocity of $v_\text{tot}\equiv\sqrt{U_{\text{LSR}}^2+V_{\text{LSR}}^2+W_{\text{LSR}}^2}\approx 70$~\kms{} is not anomalous for the thin disk, and its chemical abundances of [Fe/H] = -0.07$\pm$0.03 dex and [Mg/Fe] = +0.10$\pm$0.05 \citep{Rice2020} are more similar to younger disk stars \citep{Hayden2017}. Thus we contend that the \thisstar{} system is more likely to belong to the thin disk. As the thin disk is typically understood to have begun its formation between $\approx$8 -- 10~Gya, assigning \thisstar{} to this population allows us to set a $<$10~Gyr upper limit on the system age.

We use the $UVW$--age relationships of \citet{AlmeidaFernandes2018} and \citet{Veyette2018} to quantify the kinematic age of \thisstar{}. These relations have been calibrated to Sun-like thin disk stars in the solar neighbourhood with ages known from isochrones, the two relationships differing in their specific selection of calibrators \citep{Veyette2018}. In this case the two relations return ages for \thisstarA{}, 14.7$^{+6.1}_{-5.7}$~Gyr and 14.1$^{+3.4}_{-3.6}$~Gyr respectively, which appear implausibly high for a thin disk star. However, the lower end of the age distributions reach into plausible $<$10~Gyr ages, with 2$\sigma$ lower limits of $>$5.8~Gyr and $>$8.1~Gyr respectively. We therefore cautiously conclude that the kinematic evidence suggests a total age between $\approx$6 -- 10~Gyr for the \thisstar{} system.


In order to constrain the fundamental properties of \thisstar{}, we perform an atmosphere model fit using the MIST isochrones \citep{Dotter2016, Choi2016}, employing the same model from our previous work \citep{Venner2026}. We use space-based photometry from Tycho-2, \textit{Gaia}~DR3, 2MASS, and WISE, adopting spectroscopic priors on the effective temperature and [Fe/H] based on the values from \citet{Rice2020}. We assume an upper limit on the stellar age of $<$10~Gyr based on its membership in the thin disk as discussed above, but we do not otherwise impose a prior based on the $UVW$-age relations on account of their strong skew towards higher ages in this case. The fundamental properties from the isochrone model are recorded in Table~\ref{tab:star}. We find that \thisstarA{} has a stellar mass $M_*$ of 0.82$\pm$0.02~$M_\odot$ and a radius $R_*$ of 0.81$\pm$0.03~$R_\odot$, which conform well with expectations for a main-sequence K0 star.

\subsection{NIRC2 Imaging Analysis} \label{subsec:NIRC2_analysis}

\begin{figure}[t!]
\centering
\includegraphics[width=0.48\textwidth]{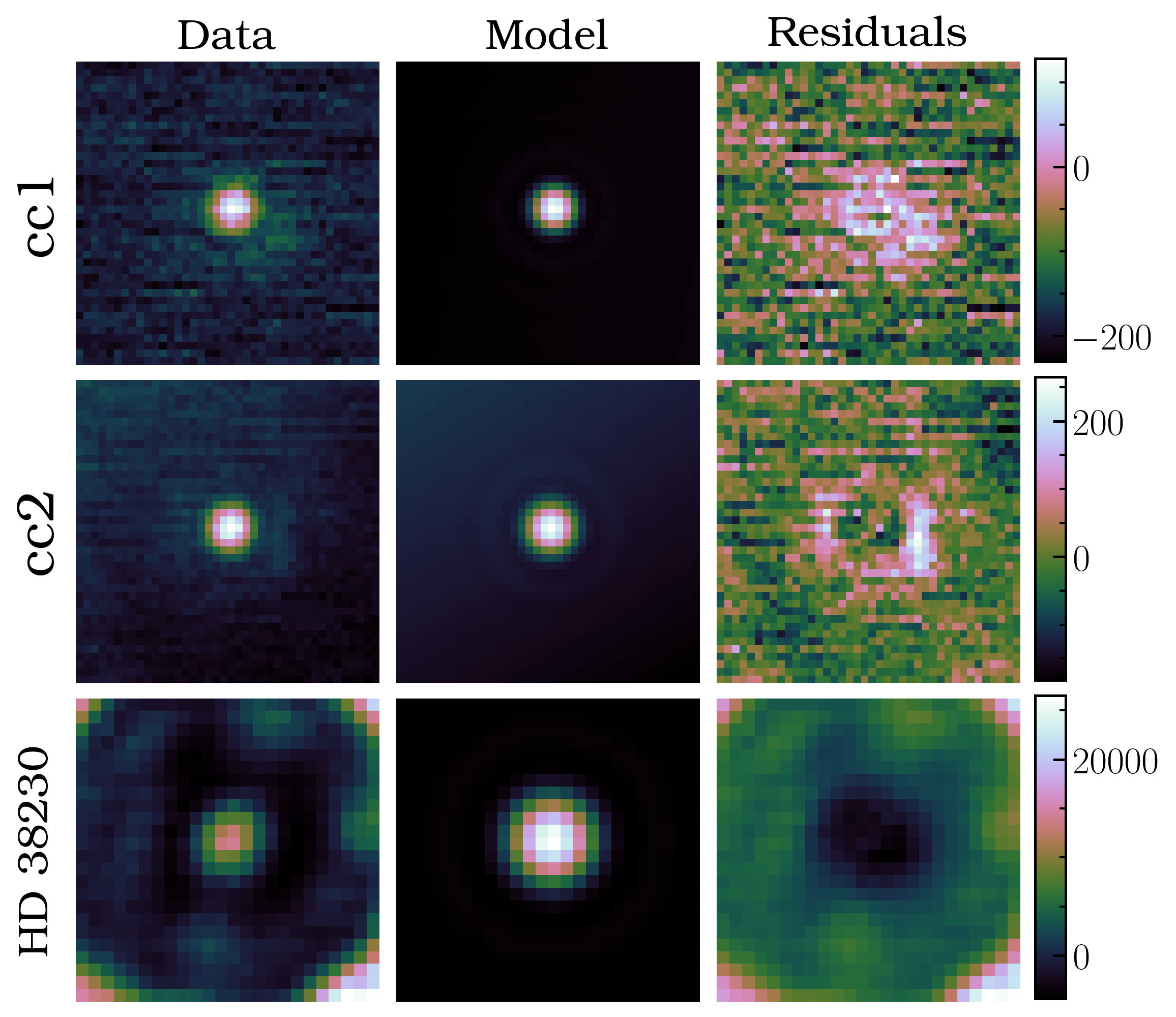}
\caption{\small{Data, model, and residual map of fits to CC1, CC2, and HD 38230 (behind the slightly-transmissive Lyot mask) for one image in the 2011 epoch.}}
\label{fig:imaging-residuals}
\end{figure}

To prepare the images for analysis, we linearized the intensity of each science and calibration image in python using the methodology of the IDL task \texttt{linearize\textunderscore nirc2.pro}
\citep{Metchev2009YoungSolarAnalogs}, then dark-subtracted and flat-fielded science frames in the standard manner. We corrected bad pixels as in \citet{Kraus2016ImpactOfStellarMult} by replacing with the median of surrounding pixels. We aligned each image in an epoch with the first image of that epoch via cross correlation through the \texttt{image-registration} python package\footnote{\url{https://image-registration.readthedocs.io}}. Exposure times are short enough that sky rotation is negligible in each image. We adopted the plate-scale, rotation offset, and distortion solution of \citet{Yelda2010NIRC2Distortion} for the 2011~Feb~22 and 2012~Feb~02 epochs, and the solution of \citet{Service2016NIRC2Distortion} for the 2015~Sep~23 epoch (updated following the NIRC2 realignment of 2015~April~13). 

We obtained epoch relative astrometry for CC1 and CC2 following the methodology of \citet{Pearce2019GSC6214-210, Pearce2021BoyajiansStar}. We precisely measured the ($x,y$) pixel position and uncertainty of CC1, CC2, and HD 38230 in each image in each epoch by fitting a PSF model that is a sum of a narrow 2D Gaussian to fit the core, a wide 2D Gaussian to fit the wings, and a 2D plane to fit the background. We constructed the models using \texttt{astropy} \citep{astropyI, astropyII, astropyIII} and fit the model to data using its Levenberg-Marquardt least square fitting algorithm. HD 38230 is behind the Lyot coronagraph, which is slightly transmissive ($\approx$0.22\% in $K$\footnote{\url{https://www2.keck.hawaii.edu/koa/public/nirc2/nirc2_data_form.html}}), allowing the position of the star to be fit as well; however we do not assume the transmission to be uniform and applied an additional 1~mas of uncertainty added in quadrature to the uncertainty on the central star's position.

\begin{deluxetable*}{lcccccrrrr}[ht!]
\tablecaption{Summary of imaging observations and relative astrometry for the two companion candidates.}
\label{tab:imaging}
\tablehead{
\colhead{Instrument} & \colhead{BJD} & \colhead{Filter} & \colhead{N$_{\rm images}$} & \colhead{N$_{\rm coadds}$} & \colhead{t$_\text{int}$} & \colhead{Separation} & \colhead{Position angle}  & \colhead{Separation} & \colhead{Position angle} \\
\colhead{} & \colhead{} & \colhead{} & \colhead{}& \colhead{} & \colhead{(sec)} & \colhead{(mas)} & \colhead{(deg)} & \colhead{(mas)} & \colhead{(deg)} } 
\startdata
\hline
& & & & & & \multicolumn{2}{c}{CC1 (\thisstarB{})} & \multicolumn{2}{c}{CC2 (background star)}\\
\hline
NIRC2 & 2455614.8 & $K^\prime$ & 40 & 10 & 2 & 4750.1$\pm$4.5 & 219.034$\pm$0.075 & 1467.7$\pm$4.8 & 23.68$\pm$0.14\\
NIRC2 & 2455959.8 & $K^\prime$ & 120 & 10 & 2 & 4742.3$\pm$2.1 & 218.808$\pm$0.029 & 1829.9$\pm$3.6 & 3.69$\pm$0.10\\
NIRC2 & 2457290.1 & $K_s$ & 41 & 40 & 1 & 4725.3$\pm$1.6 & 217.765$\pm$0.026 & 4061.2$\pm$3.4 & 334.75$\pm$0.04\\
\textit{Gaia} & 2457388.5 & $G$ & -- & -- & -- & 4720.4($\pm$5) & 217.55($\pm$0.08) & 4222.3($\pm$5) & 334.50($\pm$0.08) \\
ShaneAO * & 2457408.7 & $K_s$ & -- & -- & -- & 4780($\pm$50) & 217.6($\pm$0.5) & 4260($\pm$50) & 334.5($\pm$0.5) \\
PHARO * & 2457682.0 & $K_s$ & -- & -- & -- & 4690($\pm$50) & 218.2($\pm$0.5) & 4750($\pm$50) & 332.7($\pm$0.5) \\
ShaneAO * & 2457770.7 & $K_s$ & -- & -- & -- & 4760($\pm$30) & 217.6($\pm$0.5) & 4930($\pm$30) & 332.0($\pm$0.5) \\
\hline
\enddata
\tablecomments{Uncertainties in brackets are assumed for the orbital model. (*) Observations from \citet{Hirsch2021}. For the ShaneAO observations, we have applied an angular rotation of +1.9$\degree$ (see text).}
\end{deluxetable*}

Figure \ref{fig:imaging-residuals} shows the model fit to all three objects in one of the 2011 images. We applied the relevant distortion solutions (\citealt{Yelda2010NIRC2Distortion} for 2011 and 2012, \citealt{Service2016NIRC2Distortion} for 2015) to the ($x,y$) subpixel positions for all three objects in each image.
We converted the projected separations to mas using the relevant pixel scale (2011, 2012: 9.952$\pm$0.002$\pm$0.001 mas pixel$^{-1}$; 2015: 9.971$\pm$0.004$\pm$0.001 mas pixel$^{-1}$).
We convert from the image angle $\theta$ to position angle east of north (P.A.) by: P.A.$=\theta + \texttt{parang}+ \texttt{rotpposn} - \texttt{el} - \texttt{instangl} - \textrm{offset}$, where each term represents the corresponding image header keyword, and offset refers to the relevant distortion solution offset (2011, 2012: 0.252$\pm$0.009$^{\circ}$; 2015: 0.262$\pm$0.020$\pm$0.002$^\circ$). The uncertainty in the resulting astrometry encompasses standard deviation in image ($x,y$) positions from fit within an epoch, 1 mas uncertainty in distortion solution, uncertainty in pixel scale, uncertainty in rotation offset (all from the relevant solutions) and an additional 2 mas uncertainty in the star's position behind the Lyot mask \citep{Konopacky2016HR8799astr}. We used a Monte Carlo approach to propagate the uncertainties in R.A. and declination offsets into separation and P.A. uncertainties. The final values for the NIRC2 relative astrometry of CC1 and CC2 are reported in Table~\ref{tab:imaging}, along with archival values.

\subsection{Orbital Model} \label{subsec:orbit_model}

We fit the orbit of \thisstarB{} using \texttt{orvara} \citep{orvara}, a Bayesian orbit-fitting framework that jointly models radial velocities, absolute astrometry from the HGCA \citep[][]{Brandt2018,Brandt2021}, and relative astrometry from direct imaging. \texttt{orvara} has previously been successfully used to study the orbits of several other Sirius-like systems \citep[e.g.][]{Zhang2023}.
\texttt{orvara} parameterises orbits in terms of semi-major axis $a$, eccentricity $e$, inclination $i$, longitude of the ascending node $\Omega$, argument of periastron $\omega$ for the secondary orbit, mean longitude at a reference epoch, the parallax $\varpi$, and the component masses $M_A$ and $M_B$. At each MCMC step, the absolute astrometry is processed using \texttt{htof} \citep{htof}, which generates synthetic epoch astrometry from the Hipparcos and \textit{Gaia} observation epochs and scan angles and computes predicted proper motions that are compared to the HGCA values.

We adopt the standard \texttt{orvara} priors: a log-flat prior on semi-major axis, a log-flat prior on companion mass, a geometric prior on inclination ($p(i) \propto \sin i$), and uniform priors on eccentricity, argument of periastron, longitude of the ascending node, and mean anomaly at the reference epoch. The primary mass is drawn from a Gaussian prior $M_A$ = 0.82$\pm$0.02~$M_\odot$, and the parallax of $\varpi$ = 47.855$\pm$0.020~mas from \textit{Gaia}~DR3 (see Table~\ref{tab:star}). We fit for an RV zero-point offset and a jitter term added in quadrature to the reported uncertainties.

Our input data consist of seven epochs of relative astrometry from Keck/NIRC2, Lick/ShaneAO, and Palomar/PHARO spanning 2011--2017 (Section~\ref{subsec:imaging}; Table~\ref{tab:imaging}), the Keck/HIRES radial velocities for \thisstarA{} spanning 1996--2022 (Section~\ref{subsec:RV}), and absolute astrometry from the HGCA (Section~\ref{subsec:astrometry}). The internal uncertainties on the relative astrometry from \textit{Gaia} are extremely small ($\ll$1~mas), but we adopt a notional separation uncertainty of 5~mas (0.08$\degree$ in position angle) to aid convergence of the orbital fit \citep[see][]{Li2021}. In early testing, we observed that the literature ShaneAO and PHARO relative astrometry do not agree well with our NIRC2 and \textit{Gaia} data, presumably due to differences in the reference frame used by \citet{Hirsch2021}; to avoid adversely affecting the orbital solution, we therefore arbitrarily increase the uncertainties on these points to 30-50~mas and 0.5$\degree$, as required to bring them into $\approx$1$\sigma$ agreement with our best-fit solution.

We use a parallel-tempered MCMC with \texttt{ptemcee} \citep{ForemanMackey2013, Vousden2016}. We run 20 temperatures; for each temperature we use 100 walkers with 200,000 steps per walker, and each walker is thinned by a factor of 200. We use the coldest chain for statistical inference. We conservatively discard the first half of all steps as burn-in. Convergence is checked individually for each parameter by inspecting trace plots.

\section{Results} \label{sec:results}

\subsection{CC1 is a Bound Companion (HD 38230 B), CC2 is not}

\begin{figure}[t!]
\centering
\includegraphics[width=\columnwidth]{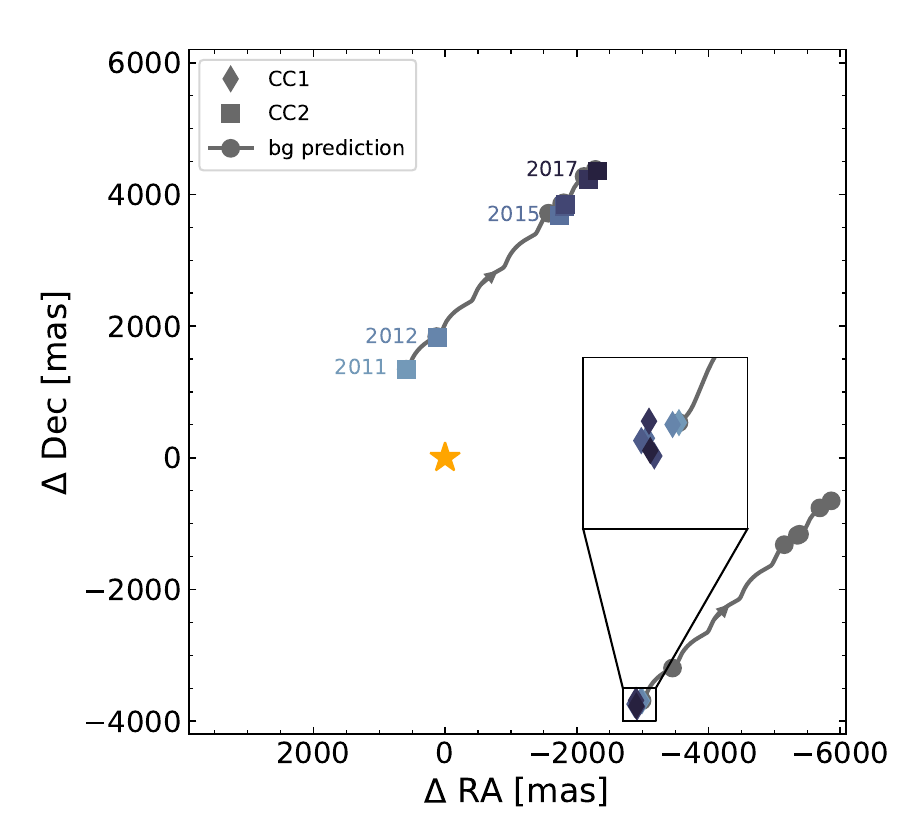}
\includegraphics[width=\columnwidth]{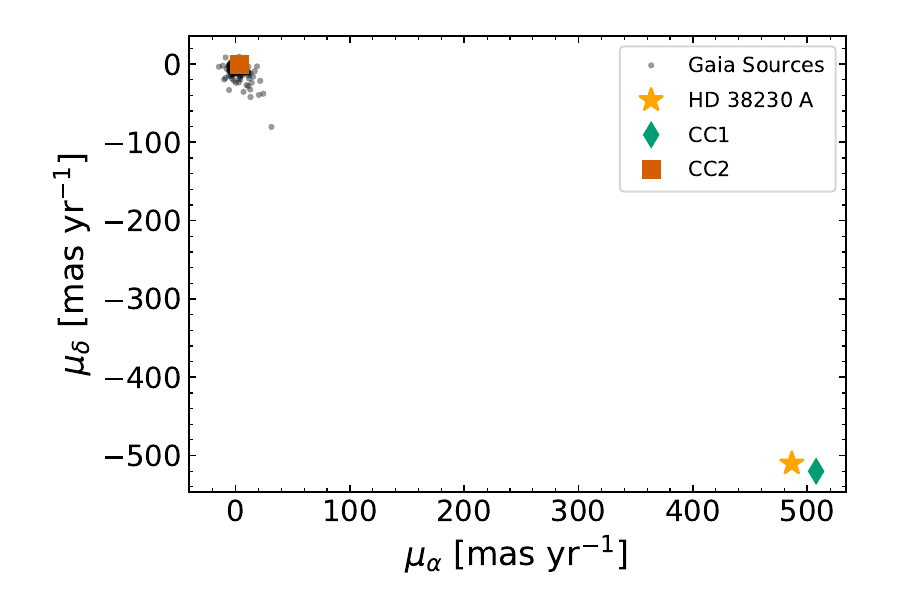}
\caption{\small{(\textit{Top}) Relative astrometry for CC1 (diamonds) and CC2 (squares) over time. HD 38230 is located at [0,0] and marked by an orange star. Zero-motion model tracks beginning from the 2011 epoch are shown for both companion candidates.
CC2 adheres closely to the stationary background track, indicating it is not associated with \thisstarA{}, whereas CC1 shows comparatively little relative motion (highlighted inset) inconsistent with a background star.
(\textit{Bottom}) Comparison of the absolute proper motions of \thisstarA{}, CC1, and CC2. We include the 2000 \textit{Gaia}~DR3 sources nearest to \thisstarA{} as grey dots to demonstrate the typical motion of background stars. The proper motion of CC2 overlaps with background objects, whereas CC1 is uniquely close to \thisstarA{}. This is consistent with CC1 being a gravitationally bound companion, \thisstarB{}.
}}
\label{fig:propermotions}
\end{figure}

Table~\ref{tab:imaging} summarises the relative astrometry from imaging. We visualise the relative and absolute motion of the two companion candidates implied by the NIRC2 observations in Figure~\ref{fig:propermotions}. It is immediately apparent that the object identified as CC1 shows comparatively limited relative motion relative to \thisstarA{}, whereas CC2 shows strong relative motion that aligns closely with the stationary background track.
To statistically test for bound companion status we performed a common proper motion test using the \texttt{backtracks} package \citep{william_o_balmer_2025_Backtracks}, which compares observed relative motion to the motion that would be observed if the object were a stationary (zero absolute motion) background object. 
CC2 agrees well with the stationary background object track, with $\chi^2_r$ = 7.96. This independently validates the background interpretation favoured by \textit{Gaia}~DR3 astrometric solution for this star, despite its high RUWE, which has a low parallax of $\varpi$ = 3.14$\pm$0.22~mas and proper motions of $\mu_\alpha,\mu_\delta$ = [+3.21$\pm$0.28, -0.71$\pm$0.15]~\masyr{} (Section~\ref{subsec:Gaia}). We conclude that CC2 is not associated with the \thisstar{} system and lies far in the background.

In contrast, CC1 is highly inconsistent with the stationary background object hypothesis. The lower panel of Figure~\ref{fig:propermotions} shows that relative to the proper motion of \thisstarA{}, CC1 is close to the primary star, whereas CC2 is consistent with background \textit{Gaia} sources. Using \texttt{backtracks}, we find that the stationary object track has a $\chi^2_r$ = 843068.5. \texttt{backtracks} includes functionality to fit a helical track to non-zero absolute proper motion for unassociated background objects that are not stationary. The attempted fit for helical motion did not converge, showing that CC1 is inconsistent with being a background object.

Across the 4.5-yr time span of NIRC2 observations, we observe that CC1 shows a relative proper motion of [+21.3, -10.0]~\masyr{},\footnote{Given the absolute proper motion of \thisstarA{} in \textit{Gaia}~DR3 (Table~\ref{tab:star}), this corresponds to an absolute proper motion of $\mu$ = [+507.7, -520.1]~\masyr{} for \thisstarB{}.} equivalent to a net tangential velocity of 2.3$\pm$0.1 km s$^{-1}$ relative to \thisstarA{} at the distance to the system. Given the projected separation of CC1 ($\approx$100~AU) and a total system mass $\approx$1.5~M$_\odot$ (the mass of \thisstarA{} plus the typical white dwarf mass of 0.6~$M_\odot$; \citealt{Tremblay2016}), a face-on circular orbit would have a velocity of 3.7~km s$^{-1}$. This means that the observed tangential velocity of CC1 is compatible with a Keplerian two-body orbit.

Previously, \citet{Hirsch2021} identified both CC1 and CC2 as background objects (Section~\ref{subsec:lit_imaging}). While we agree for CC2, this appears to be erroneous in the case of CC1. This conclusion is strengthened by the inclusion of NIRC2 observations, which extends the total time baseline for the relative astrometry from $\sim$1~yr to $\sim$6~yr (Table~\ref{tab:imaging}). We therefore conclude that CC1 is consistent with a gravitationally bound companion, which we henceforth refer to as \thisstarB{}, while CC2 is an unrelated background star.

\subsection{HD 38230 B is a White Dwarf}

\begin{figure*}[ht!]
\centering
\includegraphics[width=\textwidth]{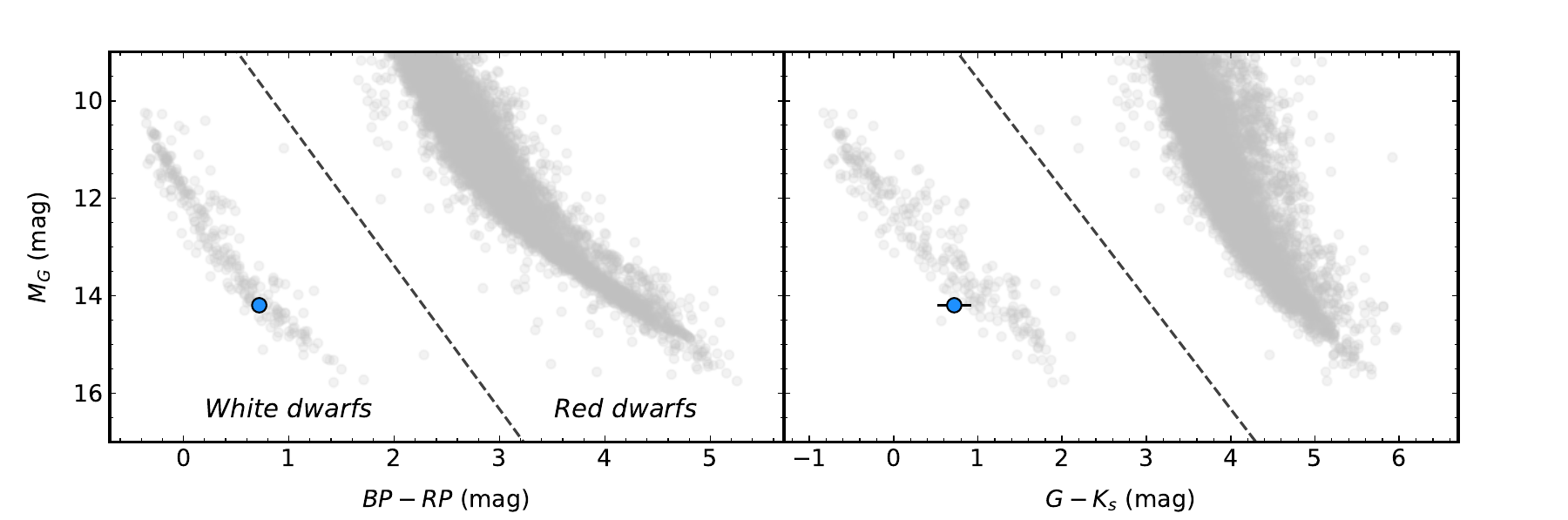}
\caption{The location of \thisstarB{} on colour-magnitude diagrams of local stars. We have constructed two colour-magnitude diagrams for stars within 40~pc based on the \textit{Gaia}~EDR3 Catalogue of Nearby Stars \citep{GCNS} comparing \textit{Gaia} BP-RP versus $M_G$ (left) and $G$-$K_s$ versus $M_G$ (right), covering the sum of available photometric observations for the companion. In both cases \thisstarB{} falls on the white dwarf sequence, unambiguously excluding any possibility it is a main sequence star; furthermore, its location on the white dwarf sequence suggests that it is relatively cool ($\approx$6000~K).
}
\label{fig:CMD}
\end{figure*}

Having demonstrated that \thisstarB{} is a gravitationally bound companion to the K-dwarf \thisstarA{}, we next investigate its physical nature. Given the system distance of $D$ = 20.880$\pm$0.009~pc, the distance modulus is $\mu=$~1.60~mag. Given $G$ = 15.79~mag in \textit{Gaia}~DR3 the absolute $G$ magnitude is therefore 14.19~mag, compatible with either a late M-dwarf or a white dwarf. The \textit{Gaia} BP-RP colour of 0.72 suggests the latter, but this is not necessarily reliable due to flux contamination from the much brighter primary star (Section~\ref{subsec:Gaia}; see also \citealt{Venner2025}).

As a result, the infrared photometry is decisive for the white dwarf interpretation.\footnote{This is similar to the discovery of $\epsilon$~Ret~B, which was identified as a probable white dwarf by \citet{Chauvin2006} based solely on its $V$, $J$ and $K$-band photometry, which was then subsequently confirmed via spectroscopy \citep{Mugrauer2007, Chauvin2007}.} \citep{Hirsch2021} reported a $K_s$ contrast of 9.80$\pm$0.20 between \thisstarA{} and B. In 2MASS the observed $K_s$ magnitude (overwhelmingly dominated by flux from A) is 5.353$\pm$0.020~mag, which results in an apparent magnitude of 15.15$\pm$0.20~mag and an absolute magnitude of 13.55$\pm$0.20~mag for \thisstarB{}.

To place the photometry of \thisstarB{} into context, we construct colour-magnitude diagrams for the bands where we have photometry for the companion. For this purpose we use the \textit{Gaia}~EDR3 Catalogue of Nearby Stars \citep{GCNS}, which includes cross-matches of local stars between \textit{Gaia} and 2MASS. We select all sources with $G$, BP, RP, and $K_s$ photometry within 40~parsecs. In Figure~\ref{fig:CMD} we show the resulting colour-magnitude diagrams in BP-RP versus $M_G$, $G$-$K_s$ versus $M_G$, respectively. With a $G$-$K_s$ colour of 0.64 $\pm$ 0.20, \thisstarB{} unambiguously falls on the white dwarf sequence, as it is far too blue to be consistent with an M-dwarf; as this concords with its BP-RP of 0.72, this therefore confirms the general accuracy of the contaminated \textit{Gaia} colour photometry for this star.

While we omit to perform a formal analysis of the spectral energy distribution of \thisstarB{} on account of the limited photometric data (and total lack of spectroscopic information), we note that the comparatively red BP-RP and $G$-$K_s$ colours suggests it is relatively cool. In the \textit{Gaia} 40~pc white dwarf sample one of the closest analogues is WD 0620-402 ($M_G$ = 14.20, BP-RP = 0.67, $G$-$K_s$ = 0.83$\pm$0.17), a DZ white dwarf with an effective temperature of 6151$\pm$40~K, mass of 0.688$\pm$0.013~$M_\odot$, and cooling age of $\sim$3.5~Gyr \citep{OBrien2024}. If \thisstarB{} likewise has an effective temperature of $\approx$7000~K, this suggests it most likely became a white dwarf several Gyr ago.

\subsection{The Orbit and Mass of HD 38230 B} \label{subsec:orbit_results}

 

\begin{figure*}[ht!]
\centering
\setlength{\tabcolsep}{4pt}
\begin{tabular}{cc}
\begin{minipage}[c][7.0cm][c]{0.47\textwidth}
  \centering
  \includegraphics[width=\linewidth,height=6.7cm,keepaspectratio=false]{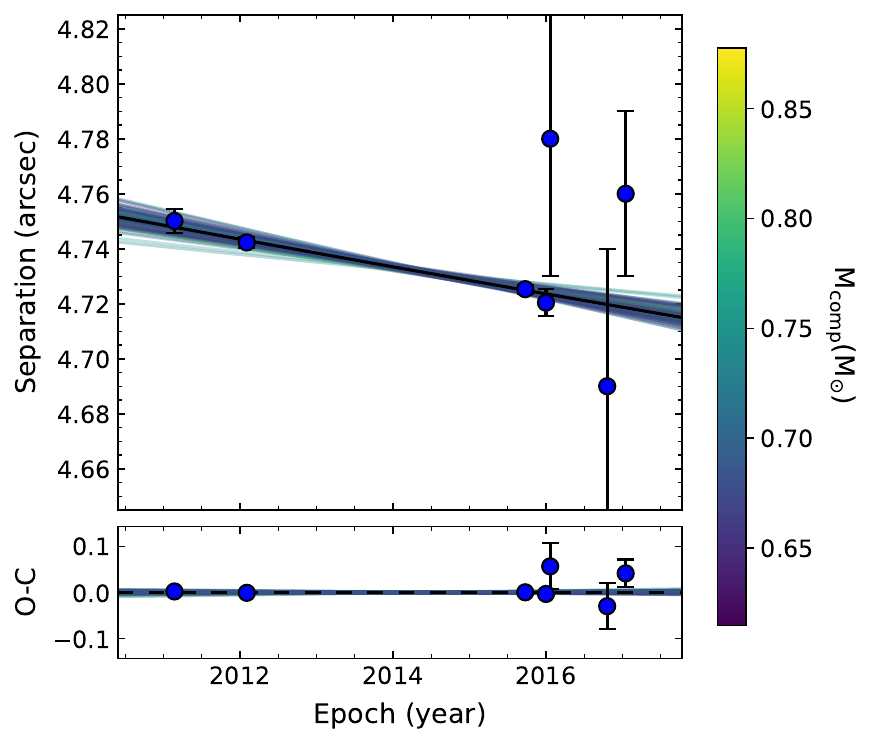}
\end{minipage} &
\begin{minipage}[c][7.0cm][c]{0.47\textwidth}
  \centering
  \includegraphics[width=\linewidth,height=6.7cm,keepaspectratio=false]{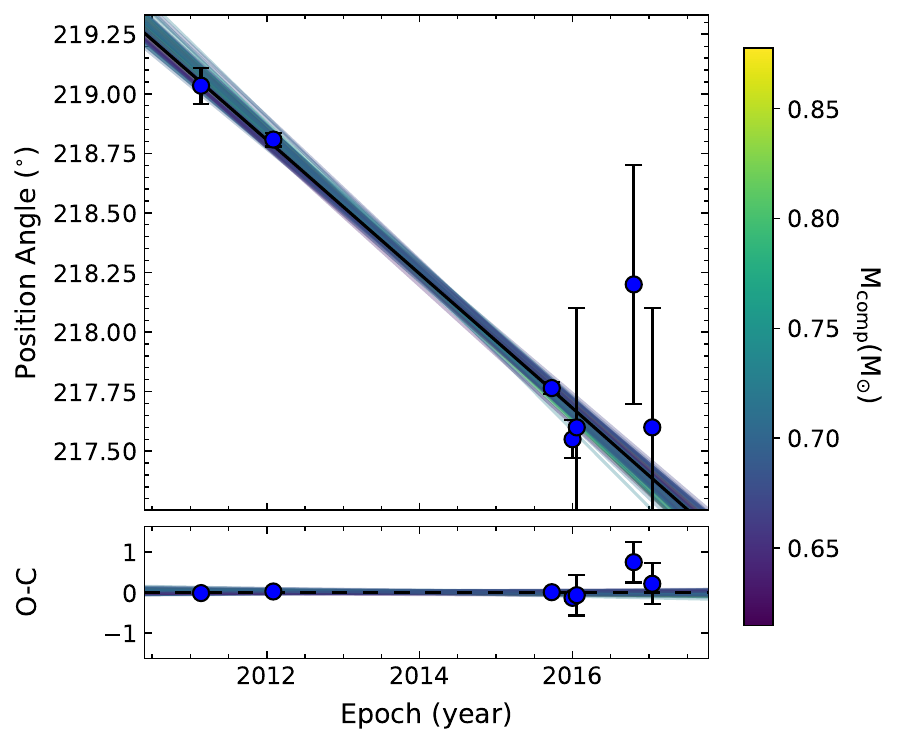}
\end{minipage} \\[6pt]
\hspace{-0.3cm}\begin{minipage}[c][7.0cm][c]{0.49\textwidth}
  \centering
  \includegraphics[width=\linewidth,height=6.7cm,keepaspectratio=false]{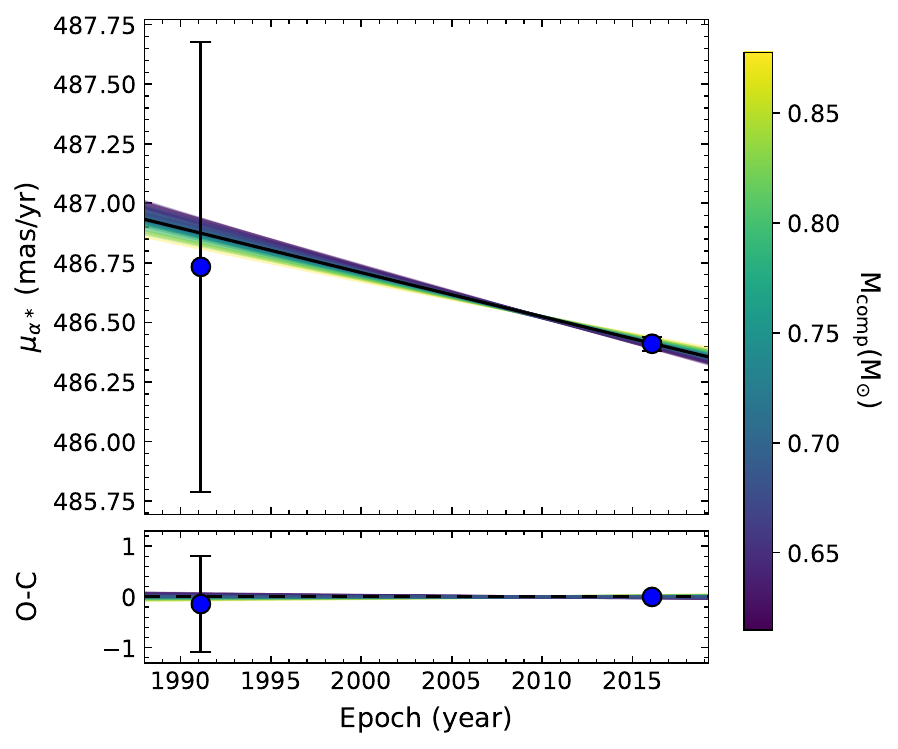}
\end{minipage} &
\hspace{-0.2cm}\begin{minipage}[c][7.0cm][c]{0.48\textwidth}
  \centering
  \includegraphics[width=\linewidth,height=6.7cm,keepaspectratio=false]{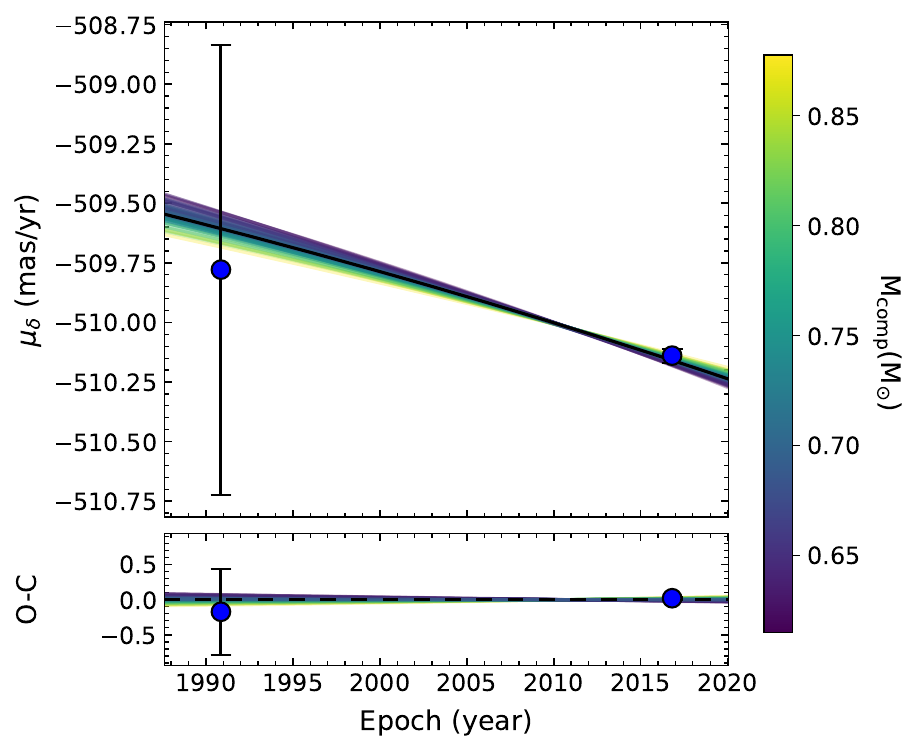}
\end{minipage} \\[6pt]
\hspace{0.1cm}\begin{minipage}[c][7.0cm][c]{0.47\textwidth}
  \centering
  \includegraphics[width=\linewidth,height=6.7cm,keepaspectratio=false]{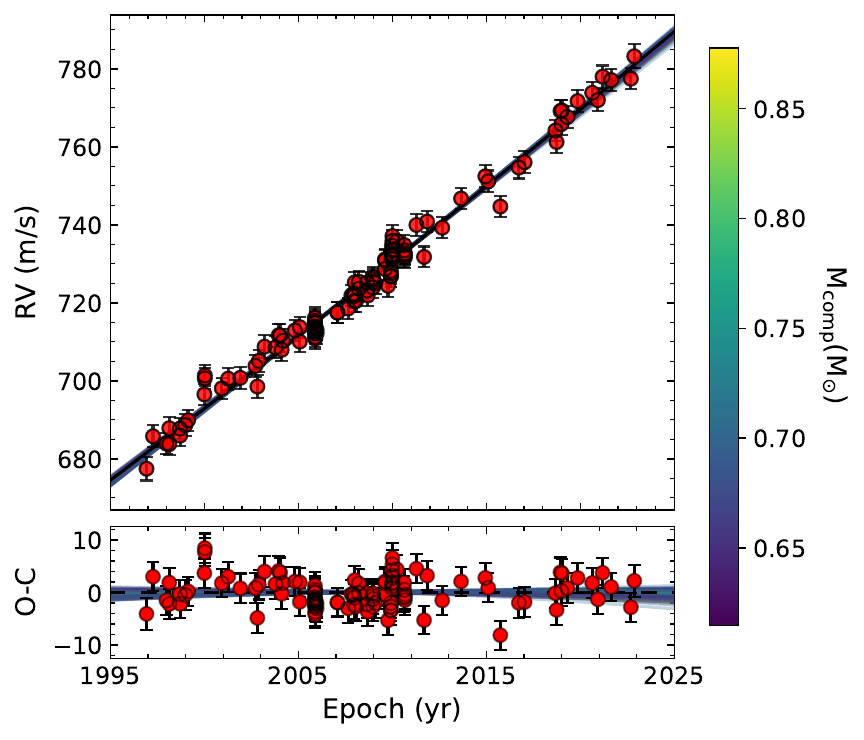}
\end{minipage} &
\multicolumn{1}{c}{\hspace{0.1cm}\begin{minipage}[c][7.0cm][c]{0.495\textwidth}
  \centering
  \includegraphics[width=\linewidth,height=6.7cm,keepaspectratio=false]{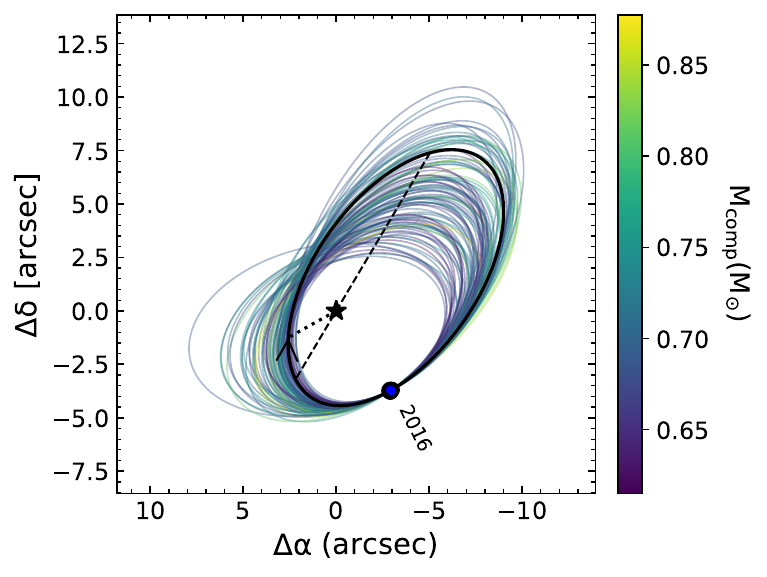}
\end{minipage}} \\
\end{tabular}
\caption{Keplerian orbit fit to \thisstarB{}. In all panels, colored curves show 100 random orbits drawn from the posterior distribution, color-coded by companion mass; the maximum-likelihood orbit is shown in black. \textit{Top row:} Projected separation (left) and position angle (right) of \thisstarB{} relative to the primary as a function of time. \textit{Middle row:} Proper motion in R.A.\ ($\mu_\alpha$, left) and Dec.\ ($\mu_\delta$, right) from the HGCA \citep{Brandt2021}, showing the \textit{Hipparcos} and \textit{Gaia} proper motions (without the time-averaged \textit{Hipparcos-Gaia} proper motion). \textit{Bottom~left:} Keck/HIRES radial velocities for \thisstarA{} with residuals after subtracting the maximum-likelihood model shown below. The RVs show a monotonic $\sim$100~\ms{} rise over $\sim$28~yr, tracing a small fraction of the $\sim$1400~yr orbit. \textit{Bottom~right:} Astrometric orbit of \thisstarB{} relative to the primary (marked by the black star at the origin). The dashed line marks the line of nodes and the dotted line connects the primary to periastron.}
\label{fig:orbit-all}
\end{figure*}

In Figure~\ref{fig:orbit-all} we show the results of our orbital model for the \thisstar{} binary. In addition, we include a \texttt{corner} plot for the corresponding posterior distributions in Appendix~\ref{appendix:corner}. The total duration of observations covers a miniscule fraction of the orbit, which we find has a period of ${1390}_{-200}^{+310}$~yr, but the combination of relative astrometry, HGCA proper motion data, and HIRES RVs strongly constrain both the radial and tangential components of the orbital motion and therefore provide robust constraints on the main orbital elements. The projected separation and position angle show the gradual relative motion of the companion over the $\sim$6~yr imaging baseline. The HGCA proper motions reveal a significant astrometric acceleration of the primary, with the \textit{Hipparcos}, \textit{Gaia}, and long-term mean proper motions all offset from one another, demonstrating the gravitational influence of the companion. The radial velocity time series shows a steady $\sim$100~\ms{} rise over the $\sim$26~yr baseline; the residuals show no significant structure, suggesting there is no strong evidence for short-period third bodies orbiting \thisstarA{}. At the scale of the sky-plane projected orbit, the observed motion in the relative astrometry is effectively negligible, demonstrating that many decades or even centuries of imaging observations would be required to capture a significant fraction of the binary orbit.

Table~\ref{tab:orbit} lists the priors and posterior distributions for all fitted and derived orbital parameters from our \texttt{orvara} model. We find that the relative orbit of \thisstarB{} has a semi-major axis of $a = 144^{+21}_{-15}$~AU, an orbital period of $P = 1390^{+310}_{-200}$~yr, an eccentricity of $e = 0.45^{+0.09}_{-0.13}$, and an inclination of $i = 125.9\degree^{+2.0\degree}_{-2.1\degree}$. The intermediate eccentricity is typical of Sirius-like binaries, whereas the semi-major axis is comparatively large among the systems with published orbits \citep[see][]{Zhang2023}.

We measure a precise dynamical mass of $M_B = 0.71^{+0.06}_{-0.05}\;M_\odot$ for \thisstarB{}. This lies near to the peak of the field white dwarf mass distribution at $\sim$0.6~$M_\odot$ \citep{Tremblay2016}, albeit slightly above average. The mass ratio $q = M_B / M_A = 0.863^{+0.074}_{-0.061}$ shows that \thisstarB{} is almost as massive as \thisstarA{}, despite its comparatively diminutive luminosity.

\begin{deluxetable}{lr}[ht!]
\tablecaption{Orbital properties of \thisstarB{}.}\label{tab:orbit}
\tablehead{
\colhead{Parameter} & \colhead{Value}}
\startdata
~~Primary mass $M_A$ ($M_\odot$) \dotfill & ${0.821}_{-0.020}^{+0.020}$ \\
~~Parallax $\varpi$ (mas) \dotfill & ${47.855}_{-0.017}^{+0.017}$ \\
\hline
~~Companion mass $M_B$ ($M_\odot$) \dotfill & ${0.71}_{-0.05}^{+0.06}$ \\
~~Semi-major axis $a$ (AU) \dotfill & ${144}_{-15}^{+21}$ \\
~~Semi-major axis $a$ (mas) \dotfill & ${6882}_{-703}^{+1005}$ \\
~~Eccentricity $e$ \dotfill & ${0.45}_{-0.13}^{+0.09}$ \\
~~Inclination $i$ (deg) \dotfill & ${125.9}_{-2.1}^{+2.0}$ \\
~~Ascending node $\Omega$ (deg) \dotfill & ${133}_{-16}^{+11}$ \\
~~Argument of periastron $\omega$ (deg) \dotfill & ${51}_{-9}^{+16}$ \\
~~Mean longitude at ref.\ epoch (deg) \dotfill & ${317}_{-42}^{+23}$ \\
~~Jitter (\ms{}) \dotfill & ${2.51}_{-0.20}^{+0.22}$ \\
\hline
~~Orbital period $P$ (yr) \dotfill & ${1390}_{-200}^{+310}$ \\
~~Time of periastron $T_0$ (BJD) \dotfill & ${2581485}_{-21669}^{+54685}$ \\
~~Mass ratio $q = M_B / M_A$ \dotfill & ${0.863}_{-0.061}^{+0.074}$ \\
\enddata
\tablecomments{Median and 68\% credible intervals. The reference epoch for the mean longitude is JD~2455197.5.}
\end{deluxetable}

\section{Discussion} \label{sec:discussion}

In this work, we have reported the discovery of a white dwarf companion to the K0~V star \thisstar{}. Lying at a distance of 21~pc, this becomes one of the nearest Sirius-like systems currently known; among the current sample, it is the 11$^\text{th}$ closest to the Sun (on which see Section~\ref{subsec:discussion_context}). This demonstrates that the existing census of white dwarfs in multiple star systems remains incomplete even at comparatively close distances - a point recently emphasised by \citet{OBrien2026}. Ground-based high-contrast imaging has played a critical role in the discovery of \thisstarB{}, as could not be identified as a bound white dwarf companion based solely on data in \textit{Gaia}~DR3, despite its substantial projected separation ($\sim$4.7\arcsec{}), since it does not have a \textit{Gaia} parallax or proper motion. This suggests it is plausible that further Sirius-like systems remain to be discovered in the solar neighbourhood, and demonstrates that a complete census of these systems cannot be compiled from \textit{Gaia} alone.

With a dynamical mass of $M_B = 0.71^{+0.06}_{-0.05}~M_\odot$ determined from our orbital model, \thisstarB{} joins the limited sample of white dwarfs with a precise mass determined independent of white dwarf atmosphere models \citep[see][]{Bowler2021, Zhang2023}. The mass of \thisstarB{} is slightly higher than the median of the field white dwarf mass distribution, which peaks strongly at $\sim$0.6~$M_\odot$ \citep{Tremblay2016}. Following the non-linear initial-final mass relation of \citet{Marigo2020}, based largely on white dwarfs in open clusters, a white dwarf with a mass of 0.71~$M_\odot$ is consistent with descent from a progenitor with initial mass of either $\approx$1.9~$M_\odot$ or $\approx$2.8~$M_\odot$, where the bimodality reflects an apparent ``kink" in the initial-final mass relation. Thus while the progenitor mass cannot be uniquely determined, the two solutions imply that \thisstarB{} was formerly an early-A or late-B type star on the main sequence, and that it was substantially more massive than the present-day primary, the K-dwarf \thisstarA{} ($M_A$ = 0.82$\pm$0.02~$M_\odot$)

A fraction of stars with white dwarf companions are known to have anomalously high abundances of neutron-capture elements, principally barium but also other elements such as strontium and yttrium, referred to as ``Barium stars" and ``(extrinsic) S stars" (\citealt{Bidelman1951, Keenan1954}; see overviews in \citealt{Jorissen2019, Escorza2019} for evolved stars and dwarfs respectively). The physical explanation for this is that this neutron capture element-enriched material has been accreted, forming via the \textit{s}-process in the interior of the asymptotic giant branch (AGB) star which precedes a present-day white dwarf in the system \citep{McClure1980, McClure1984}. We therefore inspect the \textit{s}-process abundances of \thisstarA{}. We find that in the available elemental abundance data, there does not appear to be any strong evidence for \textit{s}-process enrichment; the relative barium abundance has been measured to be $\text{[Ba/Fe]}=-0.02\pm0.07$~dex by \citet{daSilva2015}, while from the individual [M/H] abundances derived in \citet{Rice2020}, we find that $\text{[Y/Fe]}=-0.11\pm0.09$~dex. Hence \thisstarA{} is not a ``barium dwarf", despite the presence of a white dwarf companion.

\subsection{HD 38230 in Context} \label{subsec:discussion_context}

\begin{deluxetable*}{llrrllllcrl}[t!]
\label{tab:SLS_25pc}
\centering
\tablecaption{Observational properties of the known Sirius-like systems within 25~pc.}
\tablehead{\multicolumn{2}{c}{System} & \colhead{Distance} & \colhead{$V_\text{primary}$} & SpT non-WD & SpT WD & \multicolumn{2}{c}{Projected separation $\rho$} & $a$ & Confirmation & References \\
 &  & \multicolumn{1}{c}{(pc)} & (mag) &  &  & (\arcsec{}) & (AU) & (AU) & \multicolumn{1}{c}{(year)}}
\startdata
1 & Sirius & 2.639$\pm$0.010 & -1.46 & A0mA1 Va & DA2 & 10.7 & 28.2 & 19.78$\pm$0.08 & by 1922 & A, B \\
2 & Procyon & 3.509$\pm$0.009 & 0.37 & F5I V-V & DQZ6.5 & 3.80 & 13.3 & 15.12$\pm$0.04 & by 1922 & A, C \\
3 & 40 Eri & 5.010$\pm$0.003 & 4.43 & K0 V; M4.5 V & DA3 & 83 & 415 &  & by 1922 & $\gamma$, A, D \\
4 & HD 100623 & 9.551$\pm$0.002 & 5.97 & K0 V & DZ8 & 15.3 & 146 & 142$^{+38}_{-37}$ & 1996 & $\gamma$, E \\
5 & Gliese 86 & 10.760$\pm$0.006 & 6.17 & K1 V & DQ6 & 2.62 & 28.2 & 23.7$\pm$0.3 & 2005 & $\gamma$, F \\
6 & HD 147513 & 12.888$\pm$0.008 & 5.38 & G5 V & DA2 & 345 & 4450 &  & 1969 & $\gamma$, G \\
7 & HD 140901 & 15.237$\pm$0.009 & 6.01 & G7 V & DA5 & 14.5 & 221 &  & 1949 & $\gamma$, H \\
8 & HD 63077 & 15.422$\pm$0.009 & 5.36 & F9 V + K? & DC10 & 870 & 13400 &  & 1984 & $\gamma$, I \\
9 & HD 283750 & 17.387$\pm$0.014 & 8.10 & K2.5 V + M? & DC8 & 124 & 2160 &  & 1965 & $\gamma$, J \\
10 & $\epsilon$ Ret & 18.322$\pm$0.011 & 4.44 & K2 IV & DA3 & 13.0 & 240 &  & 2007 & $\gamma$, K \\
11 & \textbf{HD 38230} & 20.880$\pm$0.009 & 7.34 & K0 V & D & 4.72 & 98.6 & 144$_{-15}^{+21}$ & 2026 &  This work \\
12 & HD 218572 & 21.058$\pm$0.008 & 8.78 & K3 V & DB & 12.9 & 272 &  & 2016 & $\gamma$, L \\
13 & HD 159062 & 21.641$\pm$0.007 & 7.23 & G9 V & D & 2.70 & 58.4 & 61.9$^{+7.0}_{-7.2}$ & 2019 & $\gamma$, M \\
14 & HD 6101 & 22.00$\pm$0.04 & 8.16 & K2 V + K8V & DA6 + DC? & 1276 & 28100 &  & 2008 & $\gamma$, N \\
15 & HD 148704 & 22.531$\pm$0.022 & 7.27 & K1 V + K & D & 7.14 & 161 &  & 2024 & $\gamma$, O \\
16 & HD 34865 & 23.746$\pm$0.008 & 8.72 & K3.5 V & D & 3.65 & 86.7 &  & 2020 & $\gamma$, P \\
17 & HD 166435 & 24.396$\pm$0.010 & 6.83 & G1 V & DA2 & 29.2 & 712 &  & 2018 & $\gamma$, Q \\
\enddata
\tablecomments{~$\gamma$ = \citet{GaiaDR3, BailerJones2021}, A = \citet{Holberg2009}, B = \citet{Bond2017}, C = \citet{Provencal2002, Bond2015}, D~=~\citet{Bond2017.40Eri}, E = \citet{Hawley1996, Henry2002, An2025, OBrien2026.UV}, F = \citet{Mugrauer2005, Farihi2013, Zeng2022}, G = \citet{Alexander1969}, H = \citet{Luyten1949, Luyten1952}, I = \citet{Kunkel1984, Hartkopf2012}, J~=~\citet{Giclas1959, Eggen1965, Tokovinin1990}, K = \citet{Chauvin2006, Mugrauer2007, Farihi2011}, L = \citet{Holberg2016, Subasavage2017}, M = \citet{Hirsch2019, Bowler2021, orvara}, N~=~\citet{Maxted2000, Makarov2008}, O~=~\citet{Fuhrmann2017, Golovin2024}, P = \citet{Bonavita2020}, Q = \citet{Scholz2018}.
}
\end{deluxetable*}

To place the \thisstar{} system in the full context of the nearest Sirius-like systems to the Sun, we assemble a new compilation of all known Sirius-like systems within 25~pc. This builds upon \citet{Holberg2013}, incorporating new discoveries along with one previously known but missed system \citep[HD~6101;][]{Makarov2008}. This compilation is presented in Table~\ref{tab:SLS_25pc}. Here we focus on primary observational properties such as spectral type, primary $V$-band magnitude, angular projected separation between the stellar components, but we also incorporate semi-major axes for the systems where the WD orbit has been precisely constrained. Where possible we use distances from \citet{BailerJones2021}, which is based upon parallaxes from \textit{Gaia}~EDR3; we use the distance measurements for the primary components except where the astrometric precision is negatively affected by orbital motion (mainly due to unresolved stellar companions to the primary stars), in which case we use distances from the white dwarf parallaxes.

We tabulate the year of ``confirmation" for the white dwarf components in these systems, as in the year by which each system could be recognised as Sirius-like on the basis of the criteria of a white dwarf component in a bound multi-stellar system with a non-degenerate primary with spectral type earlier than $>$M. This is not necessarily the same as the year where the white dwarf component was discovered, as for the extreme case of HD~100623~B (WD~1132-325), which was discovered in \citet{vanBiesbroeck1961} but not recognised as a white dwarf until spectra were acquired by \citet{Hawley1996} and, independently, \citet{Henry2002}, resulting in a time lag between discovery and ``confirmation" of 35 (or 41) years.\footnote{A further 30 (26) years were required for the discovery that this white dwarf is in fact metal-polluted \citep{OBrien2026.UV}.}

We omit systems where flux from the proposed white dwarf component has not been positively detected such as $\nu$~Oct \citep[22.54$\pm$0.21~pc;][]{Cheng2025}, as without a direct detection the nature of the companion cannot be demonstrated beyond reproach. We also choose to exclude the single known short-period post-mass transfer system within 25~pc, namely Regulus \citep[$\alpha$~Leo, 24.3$\pm$0.2~pc;][]{Gies2008, Gies2020, Rappaport2009} where binary interactions have strongly influenced the evolution of the (pre-)white dwarf component, as these systems entail substantially different physical processes in their past history that cannot be easily compared to the other systems considered here. For Sirius~B, Procyon~B, and 40~Eri~B, which were discovered prior to the development of the modern physical understanding of white dwarf stars in the 1920s, the concept of ``confirmation" employed here is not relevant and we merely note that these objects were known prior to the coining of the specific term ``white dwarf" in 1922 \citep[for further historical detail see][]{Holberg2009}. Meanwhile, for the systems where the white dwarfs were first detected in \textit{Gaia}~DR2 \citep{GaiaDR2}, we adopt the first subsequent publication investigating the white dwarf component as the date of confirmation as a Sirius-like system.

In total, we identify 17 confirmed Sirius-like systems within 25~pc in Table~\ref{tab:SLS_25pc}. The distribution of primary spectral types is 1 A-dwarf (Sirius), 2 F-dwarfs, 4 G-dwarfs, 9 K-dwarfs, and 1 K-subgiant ($\epsilon$~Ret). In this sense \thisstar{}, with its K0~V primary, is typical. Five systems are triples or higher-order multiple systems with more than one non-degenerate component, although in most cases their spectral types have not been determined with certainty.

Regarding the spectral types of the white dwarf components, 7 are DA (hydrogen-dominated spectra), 1 is DB (helium-dominated), 2 are DQ (carbon-dominated), and 2 are DC (featureless spectra). EGGR~7, the white dwarf companion to HD~6101, is a short-period spectroscopic binary with an undetected secondary -- presumably a DC WD \citep{Maxted2000, Makarov2008}. Of the 13 white dwarfs in this sample with observed spectra, only two show evidence for metal pollution; Procyon~B \citep[DQZ;][]{Provencal2002}, and as very recently demonstrated on account of an exceptionally broad ultraviolet Mg~I line, HD~100623~B \citep[DZ, formerly DC;][]{OBrien2026.UV}. 4 of the white dwarfs do not currently have observed optical spectra, including \thisstarB{}, so their atmospheric compositions are not known.

For the projected separations of the white dwarf components we cite the most recent value from the literature or the Washington Double Star Catalog \citep[WDS;][]{WDS}, which in many cases is the 2016 measurement from \textit{Gaia}~DR3. These range from 2.62\arcsec{} (Gliese~86) to 1276\arcsec{} (HD~6101) on-sky, or between 13.3~AU (Procyon) to 28100~AU (HD~6101) at the respective distances to each system. In this case, \thisstarB{} is intermediate, with a projected separation of 4.72\arcsec{} (98.6~AU). Only 6 of the 17 Sirius-like systems have published orbits, of which \thisstar{} has the largest semi-major axis at 144$_{-15}^{+21}$~AU.

\citet{Holberg2013} noted a tension between the numbers of Sirius-like systems observed within $<$20~pc, which currently includes 10 known SLS,\footnote{We revise the count of 11 in \citet{Holberg2013} downwards by one; see Appendix~\ref{appendix:gl383}.} and between 20 -- 25~pc, which at that time contained none, despite these ranges encapsulating similar total volumes of space (33510~pc$^3$ versus 31940~pc$^3$ respectively). This was interpreted as evidence for severe incompleteness of the Sirius-like system sample even at relatively close distances within the solar neighbourhood. However, since the publication of that work, the disparity has significantly reduced. Whereas no new Sirius-like systems have been confirmed within $<$20~pc since $\epsilon$~Ret in 2007 (\citealt{Mugrauer2007, Chauvin2007}, following the proposed WD identification for $\epsilon$~Ret~B in \citealt{Chauvin2006}), the addition of \thisstar{} to the 20 -- 25~pc sample brings the count to 7, all but one of which were confirmed in the last decade. Of these, four postdate the 2018 release of \textit{Gaia}~DR2, which had a momentous impact on the sample of known white dwarfs \citep{Tremblay2024}; however, only one of these systems was actually discovered based on \textit{Gaia} data \citep[HD~148704,][]{Golovin2024}. Instead, three were discovered based on ground-based high-contrast imaging \citep[HD~159062, HD~34865, and HD~38230;][and this work]{Hirsch2019, Bonavita2020}, demonstrating the significant role of such observations for the discovery of new Sirius-like systems.

The unifying observational similarity for the four most recent discoveries is that the projected separations of the white dwarf components are all within $<$10\arcsec{}. This indicates that whereas the present sample of widely-separated Sirius-like systems appears to be highly complete (on account of high white dwarf completeness from \textit{Gaia}), declining detection sensitivity towards smaller angular separations due to limits in contrast sensitivity means that \textit{Gaia} alone cannot provide the entire picture. High-contrast imaging observations will therefore be vital for assembling a complete census of white dwarfs in Sirius-like systems across the solar neighbourhood.

In this context, we emphasise that the Sirius-like nature of \thisstar{} could not be determined based on \textit{Gaia} alone. Given that the count of seven Sirius-like systems between 20 -- 25~pc remains somewhat lower than the ten known within $<$20~pc,\footnote{Given a na\"ive extrapolation from the $<$20~pc count, the expected number of systems between 20 -- 25~pc would be $N$ = 9.5. However, given the small-number statistics, the Poisson uncertainty would be $\pm$3, so this does not necessarily constitute evidence for missing systems.}, it appears entirely possible that further observations could reveal further Sirius-like systems even within this highly proximate region. Recent works have even revealed white dwarf companions to M-dwarfs within $<$10~pc \citep{OBrien2026.UV, Artigau2026}, clearly suggesting that further white dwarfs hidden in multi-star systems remain to be discovered.

\subsection{Future Observational Prospects}

Despite the exceptionally long orbital period of \thisstarB{} (${1390}_{-200}^{+310}$~yr), the wealth of long-term observational data for the system means that its orbit and mass are comparatively well-characterised. The main source of observational uncertainty is the relative astrometry, as the observing baseline is comparatively short (6~yr versus 25~yr for the \textit{Hipparcos-Gaia} astrometry and 25.8~yr for the RVs). Furthermore, the most recent imaging observation is nearly a decade old (2017~Jan~17; Table~\ref{tab:imaging}). Renewed high-contrast imaging observations of \thisstar{} therefore provides the clearest avenue for continued refinement of the binary orbit.

Regarding our present understanding of \thisstarB{}, while the precise model-independent mass of ${0.71}_{-0.05}^{+0.06}~M_\odot$ determined from our orbital model is highly valuable, further interpretation of the white dwarf's properties is hindered by our lack of knowledge on its spectrum. Spectroscopic observations will be necessary in order to determine its atmospheric composition, and will make it possible to uniquely determine its temperature, luminosity, and radius. At $G$ = 15.79, \thisstarB{} is typical in brightness for a nearby white dwarf, however observing its spectrum is challenging due to the nearby presence of the much brighter primary star (4.7\arcsec{}, $\Delta G$ = 8.7~mag). This will necessitate observations with good seeing and high flux sensitivity. This could likely be achieved with a spectrograph mounted on a 8~m-class ground-based telescope; alternatively, this could be achieved from space using HST, as previously achieved for the Sirius-like white dwarf Gliese~86~B \citep{Farihi2013}.

\section{Conclusions} \label{sec:conclusions}

In this work we have reported the discovery of a white dwarf companion to the nearby K-dwarf \thisstar{}. The companion, here designated \thisstarB{}, was first detected in Keck/NIRC2 high-contrast imaging observations, then subsequently detected by \textit{Gaia} without an astrometric solution, and was also independently detected with Lick/ShaneAO and Palomar/PHARO by \citet{Hirsch2021} but inaccurately discarded as a background object. Here we have demonstrated that \thisstarB{} is instead a gravitationally bound co-moving companion to the K-dwarf \thisstarA{} with a projected separation of $\sim$100~AU. A combination of visible and near-infrared photometry from \textit{Gaia} and ground-based imaging further allows for unambiguous identification of \thisstarB{} as a white dwarf.

Thanks to a rich observational record for \thisstarA{}, including over 25 years of precise RV observations from Keck/HIRES and long-term absolute \textit{Hipparcos-Gaia} astrometry, we have been able to uniquely determine the orbit of the binary despite its long orbital period. With $P$ = $1390^{+310}_{-200}$~yr and a semi-major axis of $a$ = $144^{+21}_{-15}$~AU, \thisstar{} is one of the most widely-separated Sirius-like systems with a constrained orbit. Our orbital model provides a precise dynamical mass of $0.71^{+0.06}_{-0.05}~M_\odot$ for \thisstarB{}, independent of white dwarf atmosphere models, placing it near to but slightly above the $\sim$0.6~$M_\odot$ median mass of local white dwarfs \citep{Tremblay2016}. Further understanding of the physical properties of \thisstarB{} will depend on observation of its spectrum, which will be challenging due to the nearby presence of the much brighter primary star.

At a distance of 20.880$\pm$0.009~pc, \thisstarA{} becomes the 11$^\text{th}$ nearest ``Sirius-like system" known to date, highlighting that the census of white dwarfs in multiple star systems remains incomplete even within the solar neighbourhood. Assembling a complete local sample of white dwarfs will therefore depend on the discovery of further Sirius-like systems specifically via high-contrast imaging observations of nearby solar-type stars.

\begin{acknowledgments}

The authors wish to recognize and acknowledge the very significant cultural role and reverence that the summit of Maunakea has always had within the indigenous Hawaiian community. We are most fortunate to have the opportunity to conduct observations from this mountain.

We thank the anonymous referee, and Alex Golovin (Landessternwarte Heidelberg), for their comments that have helped to improve this work.
L.A.P.~acknowledges research support from the University of Michigan through the ELT Fellowship Program.

This research has made use of the SIMBAD database and VizieR catalogue access tool, operated at CDS, Strasbourg, France. This research has made use of NASA's Astrophysics Data System.
This work has made use of data from the European Space Agency (ESA) mission {\it Gaia} (\url{https://www.cosmos.esa.int/gaia}), processed by the {\it Gaia} Data Processing and Analysis Consortium (DPAC, \url{https://www.cosmos.esa.int/web/gaia/dpac/consortium}). Funding for the DPAC has been provided by national institutions, in particular the institutions participating in the {\it Gaia} Multilateral Agreement.

\end{acknowledgments}

\vspace{5mm}
\begin{contribution}

AV identified the target of interest in archival data, coordinated the analysis, and prepared the main part of the manuscript. LAP led the reduction and analysis of the NIRC2 data. QA led the analysis of the binary orbit.
CF independently identified the target, and coordinated a separate analysis of the system with ECM and KF; the union of these independent analyses enriched this work.
JRC and HF acquired the NIRC2 observations of the target.
TDB advised on the interpretation of the system.
All co-authors contributed suggestions and discussions that served to improve this work.




\end{contribution}

%

\vspace{5mm}
\facilities{Keck:II (ESI, NIRC2), \textit{Gaia}}


\software{\texttt{numpy} \citep{numpy}, \texttt{scipy} \citep{scipy}, \texttt{matplotlib} \citep{matplotlib}, 
\texttt{astropy} \citep{astropyI, astropyII, astropyIII}, 
\texttt{PyKOA} (\url{https://koa.ipac.caltech.edu/UserGuide/PyKOA/TAPClients.html}),
\texttt{image-registration} (\url{https://www2.keck.hawaii.edu/koa/public/nirc2/nirc2_data_form.html})
\texttt{backtracks} \citep{william_o_balmer_2025_Backtracks}, \texttt{orvara} \citep{orvara},
\texttt{htof} \citep{htof},
\texttt{ptemcee} \citep{ForemanMackey2013, Vousden2016},
\texttt{corner} \citep{corner}
}


\clearpage
\appendix

\section{Additional Figure} \label{appendix:corner}

\begin{figure*}[ht!]
\centering
\includegraphics[width=0.85\textwidth]{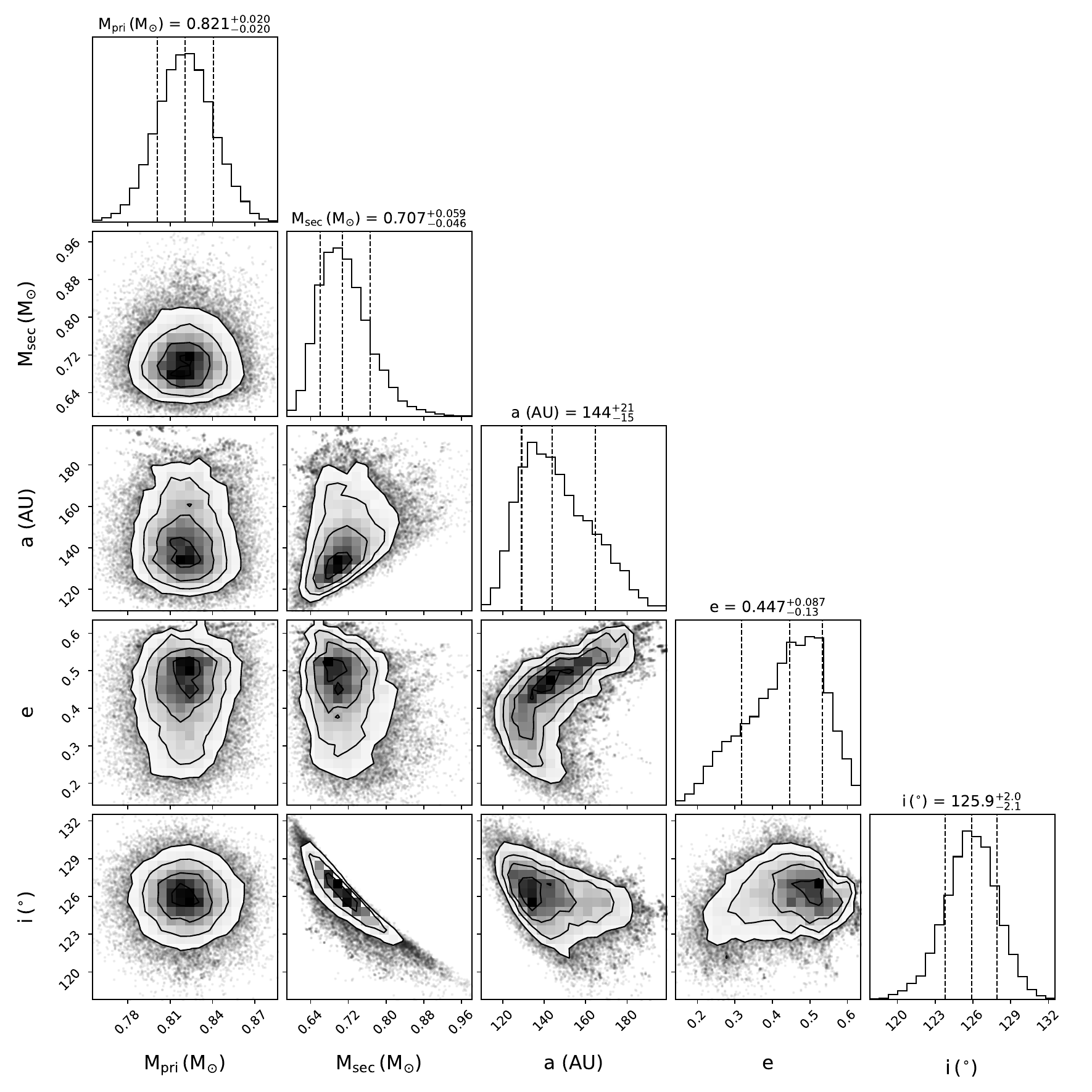}
\caption{Corner plot of the posterior distributions for the primary mass $M_A$, companion mass $M_B$, semi-major axis $a$, eccentricity $e$, and inclination $i$ of HD~38230~AB. Dashed vertical lines mark the median and 68\% credible intervals. Contours show the 1$\sigma$ and 2$\sigma$ credible regions.}
\label{fig:corner}
\end{figure*}

In Figure~\ref{fig:corner} we show a \texttt{corner} plot \citep{corner} of the posteriors from our orbital model (Section~\ref{subsec:orbit_model}, \ref{subsec:orbit_results}). This shows the correlations between the primary mass $M_A$, the secondary mass $M_B$, the semi-major axis $a$, the eccentricity $e$, and the orbital inclination $i$.

\section{On the Gliese 383 system} \label{appendix:gl383}

Gliese~383 (HIP 49973, LHS~2231, BD-17~3088) is a nearby star ($D$ = 18~pc) with a conventionally accepted spectral type of K7~V \citep{Upgren1972}. This star shows strong proper motion variability \citep[e.g.][]{Makarov2005, Brandt2021}, suggesting it is a binary with currently unknown orbital properties. Gliese~383 lies at 400\arcsec{} separation from the white dwarf WD~1009-184 (WT~1759), first spectroscopically confirmed by \citet{Henry2002} and with a first parallax measurement in \citet{Subasavage2009}. The two were first identified as a common proper motion pair in \citet{Holberg2008}, and was thus included in the Sirius-like system tally of \citet{Holberg2013}.

The \textit{Gaia} astrometry for the pair consolidate the case for physical association between Gliese~383 and WD~1009-184, although the solution for the former are clearly affected by binarity (in \textit{Gaia}~DR3 it has a RUWE of 18.45, and the parallax, $\varpi=55.49\pm0.39$~mas, has a relatively large uncertainty). In contrast, the Gaia~DR3 solution for WD~1009-184 is well-behaved and the parallax ($\varpi=55.290\pm0.036$~mas) can thus be used to define the distance to the system \citep[$D=18.070\pm0.011$~pc;][]{BailerJones2021}.

It is thus beyond doubt that Gliese~383 and WD~1009-184 form a physically associated system (likely a triple system) that lies within the 20~pc volume. However, we do not reckon this as a Sirius-like system since it appears doubtful that Gliese~383 is a K-dwarf. In \textit{Gaia~DR3}, the star has a BP-RP colour index of 1.89, which is significantly redder than the K7~V standard 61~Cyg~B (1.72), and is instead in good agreement with the M0~V standard Gliese~270 (1.88). While the K7~V spectroscopic classification from \citet{Upgren1972} is still usually accepted, the more recent spectral type determinations from the literature are M0 \citep{Bidelman1985, Gaidos2014}, and even M1 \citep{Hejazi2020}. It thus appears much more probable that Gliese~383 is an early M-dwarf rather than a late K-dwarf. Given our adoption of the definition for ``Sirius-like systems" from \citet{Holberg2013} that excludes systems with M-dwarf primaries, we do not count the Gliese~383 system as a SLS. This reduces the count of known Sirius-like systems in the 20~pc sample from 11 in \citet{Holberg2013} to 10.


\bibliography{bib}{}

@preamble{"\newcommand{\noop}[1]{}"}

@ARTICLE{Holberg2013,
       author = {{Holberg}, J.~B. and {Oswalt}, T.~D. and {Sion}, E.~M. and {Barstow}, M.~A. and {Burleigh}, M.~R.},
        title = "{Where are all the Sirius-like binary systems?}",
      journal = {\mnras},
         year = 2013,
        month = nov,
       volume = {435},
       number = {3},
        pages = {2077-2091},
          doi = {10.1093/mnras/stt1433},
archivePrefix = {arXiv},
       eprint = {1307.8047},
 primaryClass = {astro-ph.SR},
       adsurl = {https://ui.adsabs.harvard.edu/abs/2013MNRAS.435.2077H}
}

@ARTICLE{Holberg2009,
       author = {{Holberg}, J.~B.},
        title = "{The Discovery of the Existence of White Dwarf Stars: 1862 to 1930}",
      journal = {Journal for the History of Astronomy},
         year = 2009,
        month = may,
       volume = {40},
       number = {2},
        pages = {137-154},
          doi = {10.1177/002182860904000201},
       adsurl = {https://ui.adsabs.harvard.edu/abs/2009JHA....40..137H}
}

@ARTICLE{Crepp2012a,
       author = {{Crepp}, Justin R. and others},
        title = "{The Dynamical Mass and Three-dimensional Orbit of HR7672B: A Benchmark Brown Dwarf with High Eccentricity}",
      journal = {\apj},
         year = 2012,
        month = jun,
       volume = {751},
       number = {2},
          eid = {97},
        pages = {97},
          doi = {10.1088/0004-637X/751/2/97},
archivePrefix = {arXiv},
       eprint = {1112.1725},
 primaryClass = {astro-ph.EP},
       adsurl = {https://ui.adsabs.harvard.edu/abs/2012ApJ...751...97C}
}

@ARTICLE{Crepp2013a,
       author = {{Crepp}, Justin R. and others},
        title = "{The TRENDS High-contrast Imaging Survey. II. Direct Detection of the HD 8375 Tertiary}",
      journal = {\apj},
         year = 2013,
        month = jul,
       volume = {771},
       number = {1},
          eid = {46},
        pages = {46},
          doi = {10.1088/0004-637X/771/1/46},
archivePrefix = {arXiv},
       eprint = {1210.7374},
 primaryClass = {astro-ph.EP},
       adsurl = {https://ui.adsabs.harvard.edu/abs/2013ApJ...771...46C}
}

@ARTICLE{Crepp2013b,
       author = {{Crepp}, Justin R. and others},
        title = "{The TRENDS High-contrast Imaging Survey. III. A Faint White Dwarf Companion Orbiting HD 114174}",
      journal = {\apj},
         year = 2013,
        month = sep,
       volume = {774},
       number = {1},
          eid = {1},
        pages = {1},
          doi = {10.1088/0004-637X/774/1/1},
archivePrefix = {arXiv},
       eprint = {1305.0571},
 primaryClass = {astro-ph.SR},
       adsurl = {https://ui.adsabs.harvard.edu/abs/2013ApJ...774....1C}
}

@ARTICLE{Crepp2014,
       author = {{Crepp}, Justin R. and others},
        title = "{The TRENDS High-contrast Imaging Survey. V. Discovery of an Old and Cold Benchmark T-dwarf Orbiting the Nearby G-star HD 19467}",
      journal = {\apj},
         year = 2014,
        month = jan,
       volume = {781},
       number = {1},
          eid = {29},
        pages = {29},
          doi = {10.1088/0004-637X/781/1/29},
archivePrefix = {arXiv},
       eprint = {1311.0280},
 primaryClass = {astro-ph.EP},
       adsurl = {https://ui.adsabs.harvard.edu/abs/2014ApJ...781...29C}
}

@ARTICLE{Crepp2016,
       author = {{Crepp}, Justin R. and {Gonzales}, Erica J. and {Bechter}, Eric B. and {Montet}, Benjamin T. and {Johnson}, John Asher and {Piskorz}, Danielle and {Howard}, Andrew W. and {Isaacson}, Howard},
        title = "{The TRENDS High-contrast Imaging Survey. VI. Discovery of a Mass, Age, and Metallicity Benchmark Brown Dwarf}",
      journal = {\apj},
         year = 2016,
        month = dec,
       volume = {831},
       number = {2},
          eid = {136},
        pages = {136},
          doi = {10.3847/0004-637X/831/2/136},
archivePrefix = {arXiv},
       eprint = {1604.00398},
 primaryClass = {astro-ph.SR},
       adsurl = {https://ui.adsabs.harvard.edu/abs/2016ApJ...831..136C}
}

@ARTICLE{Zurlo2013,
       author = {{Zurlo}, A. and {Vigan}, A. and {Hagelberg}, J. and {Desidera}, S. and {Chauvin}, G. and {Almenara}, J.~M. and {Biazzo}, K. and {Bonnefoy}, M. and {Carson}, J.~C. and {Covino}, E. and {Delorme}, P. and {D'Orazi}, V. and {Gratton}, R. and {Mesa}, D. and {Messina}, S. and {Moutou}, C. and {Segransan}, D. and {Turatto}, M. and {Udry}, S. and {Wildi}, F.},
        title = "{Astrophysical false positives in direct imaging for exoplanets: a white dwarf close to a rejuvenated star}",
      journal = {\aap},
         year = 2013,
        month = jun,
       volume = {554},
          eid = {A21},
        pages = {A21},
          doi = {10.1051/0004-6361/201321179},
archivePrefix = {arXiv},
       eprint = {1304.4130},
 primaryClass = {astro-ph.EP},
       adsurl = {https://ui.adsabs.harvard.edu/abs/2013A&A...554A..21Z}
}

@ARTICLE{Montet2014,
       author = {{Montet}, Benjamin T. and others},
        title = "{The TRENDS High-contrast Imaging Survey. IV. The Occurrence Rate of Giant Planets around M Dwarfs}",
      journal = {\apj},
         year = 2014,
        month = jan,
       volume = {781},
       number = {1},
          eid = {28},
        pages = {28},
          doi = {10.1088/0004-637X/781/1/28},
archivePrefix = {arXiv},
       eprint = {1307.5849},
 primaryClass = {astro-ph.EP},
       adsurl = {https://ui.adsabs.harvard.edu/abs/2014ApJ...781...28M}
}

@ARTICLE{Rodigas2016,
       author = {{Rodigas}, Timothy J. and {Bergeron}, P. and {Simon}, Am{\'e}lie and {Arriagada}, Pamela and {Faherty}, Jacqueline K. and {Anglada-Escud{\'e}}, Guillem and {Mamajek}, Eric E. and {Weinberger}, Alycia and {Butler}, R. Paul and {Males}, Jared R. and {Morzinski}, Katie and {Close}, Laird M. and {Hinz}, Philip M. and {Bailey}, Jeremy and {Carter}, Brad and {Jenkins}, James S. and {Jones}, Hugh and {O'Toole}, Simon and {Tinney}, C.~G. and {Wittenmyer}, Rob and {Debes}, John},
        title = "{MagAO Imaging of Long-period Objects (MILO). II. A Puzzling White Dwarf around the Sun-like Star HD 11112}",
      journal = {\apj},
         year = 2016,
        month = nov,
       volume = {831},
       number = {2},
          eid = {177},
        pages = {177},
          doi = {10.3847/0004-637X/831/2/177},
archivePrefix = {arXiv},
       eprint = {1609.02576},
 primaryClass = {astro-ph.EP},
       adsurl = {https://ui.adsabs.harvard.edu/abs/2016ApJ...831..177R}
}

@ARTICLE{Crepp2018,
       author = {{Crepp}, Justin R. and {Gonzales}, Erica J. and {Bowler}, Brendan P. and {Morales}, Farisa and {Stone}, Jordan and {Spalding}, Eckhart and {Vaz}, Amali and {Hinz}, Philip and {Ertel}, Steve and {Howard}, Andrew and {Isaacson}, Howard},
        title = "{The TRENDS High-contrast Imaging Survey. VII. Discovery of a Nearby Sirius-like White Dwarf System (HD 169889)}",
      journal = {\apj},
         year = 2018,
        month = sep,
       volume = {864},
       number = {1},
          eid = {42},
        pages = {42},
          doi = {10.3847/1538-4357/aad381},
archivePrefix = {arXiv},
       eprint = {1807.06012},
 primaryClass = {astro-ph.SR},
       adsurl = {https://ui.adsabs.harvard.edu/abs/2018ApJ...864...42C}
}

@ARTICLE{Hirsch2019,
       author = {{Hirsch}, Lea A. and others},
        title = "{Discovery of a White Dwarf Companion to HD 159062}",
      journal = {\apj},
         year = 2019,
        month = jun,
       volume = {878},
       number = {1},
          eid = {50},
        pages = {50},
          doi = {10.3847/1538-4357/ab1b11},
archivePrefix = {arXiv},
       eprint = {1905.06440},
 primaryClass = {astro-ph.SR},
       adsurl = {https://ui.adsabs.harvard.edu/abs/2019ApJ...878...50H}
}

@ARTICLE{Bonavita2020,
       author = {{Bonavita}, M. and {Fontanive}, C. and {Desidera}, S. and {D'Orazi}, V. and {Zurlo}, A. and {Mu{\v{z}}i{\'c}}, K. and {Biller}, B. and {Gratton}, R. and {Mesa}, D. and {Sozzetti}, A.},
        title = "{A new white dwarf companion around the {\ensuremath{\Delta}}{\ensuremath{\mu}} star GJ 3346}",
      journal = {\mnras},
         year = 2020,
        month = may,
       volume = {494},
       number = {3},
        pages = {3481-3490},
          doi = {10.1093/mnras/staa590},
archivePrefix = {arXiv},
       eprint = {2002.10467},
 primaryClass = {astro-ph.SR},
       adsurl = {https://ui.adsabs.harvard.edu/abs/2020MNRAS.494.3481B}
}

@ARTICLE{Bowler2021,
       author = {{Bowler}, Brendan P. and {Cochran}, William D. and {Endl}, Michael and {Franson}, Kyle and {Brandt}, Timothy D. and {Dupuy}, Trent J. and {MacQueen}, Phillip J. and {Kratter}, Kaitlin M. and {Mawet}, Dimitri and {Ruane}, Garreth},
        title = "{The McDonald Accelerating Stars Survey (MASS): White Dwarf Companions Accelerating the Sun-like Stars 12 Psc and HD 159062}",
      journal = {\aj},
         year = 2021,
        month = mar,
       volume = {161},
       number = {3},
          eid = {106},
        pages = {106},
          doi = {10.3847/1538-3881/abd243},
archivePrefix = {arXiv},
       eprint = {2012.04847},
 primaryClass = {astro-ph.SR},
       adsurl = {https://ui.adsabs.harvard.edu/abs/2021AJ....161..106B}
}

@ARTICLE{Zhang2023,
       author = {{Zhang}, Hengyue and {Brandt}, Timothy D. and {Kiman}, Rocio and {Venner}, Alexander and {An}, Qier and {Chen}, Minghan and {Li}, Yiting},
        title = "{Dynamical masses and ages of Sirius-like systems}",
      journal = {\mnras},
         year = 2023,
        month = sep,
       volume = {524},
       number = {1},
        pages = {695-715},
          doi = {10.1093/mnras/stad1849},
archivePrefix = {arXiv},
       eprint = {2303.08198},
 primaryClass = {astro-ph.SR},
       adsurl = {https://ui.adsabs.harvard.edu/abs/2023MNRAS.524..695Z}
}

@ARTICLE{Pearce2025,
       author = {{Pearce}, Logan A. and {Males}, Jared R. and {Haffert}, Sebastiaan Y. and {Close}, Laird M. and {Long}, Joseph D. and {McEwen}, Eden A. and {Liberman}, Joshua and {Kautz}, Maggie Y. and {Kueny}, Jay K. and {Weinberger}, Alycia J. and {Li}, Jialin and {Tonucci}, Elena and {Twitchell}, Katie and {McLeod}, Avalon L. and {Foster}, Warren B. and {Guyon}, Olivier and {Hedglen}, Alexander D. and {Van Gorkom}, Kyle and {Lumbres}, Jennifer and {Schatz}, Lauren and {Gasho}, Victor and {Morzinski}, Katie M. and {Hinz}, Phil M.},
        title = "{Five New Sirius-like White Dwarf{\textendash}Main-sequence Star Systems with MagAO-X}",
      journal = {\aj},
         year = 2025,
        month = jul,
       volume = {170},
       number = {1},
          eid = {58},
        pages = {58},
          doi = {10.3847/1538-3881/addb50},
archivePrefix = {arXiv},
       eprint = {2505.14439},
 primaryClass = {astro-ph.SR},
       adsurl = {https://ui.adsabs.harvard.edu/abs/2025AJ....170...58P}
}

@ARTICLE{Jenkins2024,
       author = {{Jenkins}, Sydney and {Vanderburg}, Andrew and {Bieryla}, Allyson and {Latham}, David W. and {Badenas-Agusti}, Mariona and {Berlind}, Perry and {Blouin}, Simon and {Buchhave}, Lars A. and {Calkins}, Michael L. and {Esquerdo}, Gilbert A. and {Via{\~n}a}, Javier},
        title = "{Absence of a correlation between white dwarf planetary accretion and primordial stellar metallicity}",
      journal = {\mnras},
         year = 2024,
        month = jul,
       volume = {532},
       number = {1},
        pages = {394-410},
          doi = {10.1093/mnras/stae1506},
archivePrefix = {arXiv},
       eprint = {2406.18638},
 primaryClass = {astro-ph.EP},
       adsurl = {https://ui.adsabs.harvard.edu/abs/2024MNRAS.532..394J}
}

@ARTICLE{Catalan2008,
       author = {{Catal{\'a}n}, S. and {Isern}, J. and {Garc{\'\i}a-Berro}, E. and {Ribas}, I. and {Allende Prieto}, C. and {Bonanos}, A.~Z.},
        title = "{The initial-final mass relationship from white dwarfs in common proper motion pairs}",
      journal = {\aap},
         year = 2008,
        month = jan,
       volume = {477},
       number = {1},
        pages = {213-221},
          doi = {10.1051/0004-6361:20078111},
archivePrefix = {arXiv},
       eprint = {0710.1542},
 primaryClass = {astro-ph},
       adsurl = {https://ui.adsabs.harvard.edu/abs/2008A&A...477..213C}
}

@ARTICLE{Zhao2012,
       author = {{Zhao}, J.~K. and {Oswalt}, T.~D. and {Willson}, L.~A. and {Wang}, Q. and {Zhao}, G.},
        title = "{The Initial-Final Mass Relation among White Dwarfs in Wide Binaries}",
      journal = {\apj},
         year = 2012,
        month = feb,
       volume = {746},
       number = {2},
          eid = {144},
        pages = {144},
          doi = {10.1088/0004-637X/746/2/144},
archivePrefix = {arXiv},
       eprint = {1112.0281},
 primaryClass = {astro-ph.SR},
       adsurl = {https://ui.adsabs.harvard.edu/abs/2012ApJ...746..144Z}
}

@ARTICLE{Barrientos2021,
       author = {{Barrientos}, Manuel and {Chanam{\'e}}, Julio},
        title = "{Improved Constraints on the Initial-to-final Mass Relation of White Dwarfs Using Wide Binaries}",
      journal = {\apj},
         year = 2021,
        month = dec,
       volume = {923},
       number = {2},
          eid = {181},
        pages = {181},
          doi = {10.3847/1538-4357/ac2f49},
archivePrefix = {arXiv},
       eprint = {2102.07790},
 primaryClass = {astro-ph.SR},
       adsurl = {https://ui.adsabs.harvard.edu/abs/2021ApJ...923..181B}
}

@ARTICLE{Fouesneau2019,
       author = {{Fouesneau}, Morgan and {Rix}, Hans-Walter and {von Hippel}, Ted and {Hogg}, David. W. and {Tian}, Haijun},
        title = "{Precise Ages of Field Stars from White Dwarf Companions}",
      journal = {\apj},
         year = 2019,
        month = jan,
       volume = {870},
       number = {1},
          eid = {9},
        pages = {9},
          doi = {10.3847/1538-4357/aaee74},
archivePrefix = {arXiv},
       eprint = {1802.06663},
 primaryClass = {astro-ph.SR},
       adsurl = {https://ui.adsabs.harvard.edu/abs/2019ApJ...870....9F}
}

@ARTICLE{Venner2023,
       author = {{Venner}, Alexander and {Blouin}, Simon and {B{\'e}dard}, Antoine and {Vanderburg}, Andrew},
        title = "{A crystallizing white dwarf in a Sirius-like quadruple system}",
      journal = {\mnras},
         year = 2023,
        month = aug,
       volume = {523},
       number = {3},
        pages = {4624-4642},
          doi = {10.1093/mnras/stad1719},
archivePrefix = {arXiv},
       eprint = {2306.03140},
 primaryClass = {astro-ph.SR},
       adsurl = {https://ui.adsabs.harvard.edu/abs/2023MNRAS.523.4624V}
}

@ARTICLE{Venner2025,
       author = {{Venner}, Alexander and {Limbach}, Mary Anne and {M{\^a}lin}, Mathilde and {Blouin}, Simon and {Boccaletti}, Anthony and {Pearce}, Logan A.},
        title = "{Serendipitous observation of a white dwarf companion to a JWST/MIRI coronagraphic calibrator}",
      journal = {\mnras},
         year = 2025,
        month = jan,
       volume = {536},
       number = {1},
        pages = {L38-L41},
          doi = {10.1093/mnrasl/slae106},
archivePrefix = {arXiv},
       eprint = {2410.11568},
 primaryClass = {astro-ph.SR},
       adsurl = {https://ui.adsabs.harvard.edu/abs/2025MNRAS.536L..38V}
}

@ARTICLE{Barrientos2024,
       author = {{Barrientos}, Manuel and {Kilic}, Mukremin and {Bergeron}, Pierre and {Blouin}, Simon and {Brown}, Warren R. and {Andrews}, Jeff J.},
        title = "{Fundamental Tests of White Dwarf Cooling Physics with Wide Binaries}",
      journal = {\apj},
         year = 2024,
        month = oct,
       volume = {973},
       number = {2},
          eid = {88},
        pages = {88},
          doi = {10.3847/1538-4357/ad6303},
archivePrefix = {arXiv},
       eprint = {2407.18763},
 primaryClass = {astro-ph.SR},
       adsurl = {https://ui.adsabs.harvard.edu/abs/2024ApJ...973...88B}
}

@ARTICLE{Tremblay2024,
       author = {{Tremblay}, Pier-Emmanuel and {B{\'e}dard}, Antoine and {O'Brien}, Mairi W. and {Munday}, James and {Elms}, Abbigail K. and {Gentillo Fusillo}, Nicola Pietro and {Sahu}, Snehalata},
        title = "{The Gaia white dwarf revolution}",
      journal = {\nar},
         year = 2024,
        month = dec,
       volume = {99},
          eid = {101705},
        pages = {101705},
          doi = {10.1016/j.newar.2024.101705},
archivePrefix = {arXiv},
       eprint = {2402.14960},
 primaryClass = {astro-ph.SR},
       adsurl = {https://ui.adsabs.harvard.edu/abs/2024NewAR..9901705T}
}

@ARTICLE{ElBadry2018,
       author = {{El-Badry}, Kareem and {Rix}, Hans-Walter},
        title = "{Imprints of white dwarf recoil in the separation distribution of Gaia wide binaries}",
      journal = {\mnras},
         year = 2018,
        month = nov,
       volume = {480},
       number = {4},
        pages = {4884-4902},
          doi = {10.1093/mnras/sty2186},
archivePrefix = {arXiv},
       eprint = {1807.06011},
 primaryClass = {astro-ph.SR},
       adsurl = {https://ui.adsabs.harvard.edu/abs/2018MNRAS.480.4884E}
}

@ARTICLE{Crepp2012b,
       author = {{Crepp}, Justin R. and others},
        title = "{The TRENDS High-contrast Imaging Survey. I. Three Benchmark M Dwarfs Orbiting Solar-type Stars}",
      journal = {\apj},
         year = 2012,
        month = dec,
       volume = {761},
       number = {1},
          eid = {39},
        pages = {39},
          doi = {10.1088/0004-637X/761/1/39},
archivePrefix = {arXiv},
       eprint = {1210.3000},
 primaryClass = {astro-ph.EP},
       adsurl = {https://ui.adsabs.harvard.edu/abs/2012ApJ...761...39C}
}

@ARTICLE{Gonzales2020,
       author = {{Gonzales}, Erica J. and {Crepp}, Justin R. and {Bechter}, Eric B. and {Wood}, Charlotte M. and {Johnson}, John Asher and {Montet}, Benjamin T. and {Isaacson}, Howard and {Howard}, Andrew W.},
        title = "{The TRENDS High-contrast Imaging Survey. VIII. Compendium of Benchmark Objects}",
      journal = {\apj},
         year = 2020,
        month = apr,
       volume = {893},
       number = {1},
          eid = {27},
        pages = {27},
          doi = {10.3847/1538-4357/ab71fb},
archivePrefix = {arXiv},
       eprint = {2010.11866},
 primaryClass = {astro-ph.SR},
       adsurl = {https://ui.adsabs.harvard.edu/abs/2020ApJ...893...27G}
}

@ARTICLE{Hirsch2021,
       author = {{Hirsch}, Lea A. and others},
        title = "{Understanding the Impacts of Stellar Companions on Planet Formation and Evolution: A Survey of Stellar and Planetary Companions within 25 pc}",
      journal = {\aj},
         year = 2021,
        month = mar,
       volume = {161},
       number = {3},
          eid = {134},
        pages = {134},
          doi = {10.3847/1538-3881/abd639},
archivePrefix = {arXiv},
       eprint = {2012.09190},
 primaryClass = {astro-ph.EP},
       adsurl = {https://ui.adsabs.harvard.edu/abs/2021AJ....161..134H}
}

@ARTICLE{Riello2021,
       author = {{Riello}, M. and {De Angeli}, F. and {Evans}, D.~W. and {Montegriffo}, P. and {Carrasco}, J.~M. and {Busso}, G. and {Palaversa}, L. and {Burgess}, P.~W. and {Diener}, C. and {Davidson}, M. and {Rowell}, N. and {Fabricius}, C. and {Jordi}, C. and {Bellazzini}, M. and {Pancino}, E. and {Harrison}, D.~L. and {Cacciari}, C. and {van Leeuwen}, F. and {Hambly}, N.~C. and {Hodgkin}, S.~T. and {Osborne}, P.~J. and {Altavilla}, G. and {Barstow}, M.~A. and {Brown}, A.~G.~A. and {Castellani}, M. and {Cowell}, S. and {De Luise}, F. and {Gilmore}, G. and {Giuffrida}, G. and {Hidalgo}, S. and {Holland}, G. and {Marinoni}, S. and {Pagani}, C. and {Piersimoni}, A.~M. and {Pulone}, L. and {Ragaini}, S. and {Rainer}, M. and {Richards}, P.~J. and {Sanna}, N. and {Walton}, N.~A. and {Weiler}, M. and {Yoldas}, A.},
        title = "{Gaia Early Data Release 3. Photometric content and validation}",
      journal = {\aap},
         year = 2021,
        month = may,
       volume = {649},
          eid = {A3},
        pages = {A3},
          doi = {10.1051/0004-6361/202039587},
archivePrefix = {arXiv},
       eprint = {2012.01916},
 primaryClass = {astro-ph.IM},
       adsurl = {https://ui.adsabs.harvard.edu/abs/2021A&A...649A...3R}
}

@ARTICLE{Butler2017,
       author = {{Butler}, R. Paul and {Vogt}, Steven S. and {Laughlin}, Gregory and {Burt}, Jennifer A. and {Rivera}, Eugenio J. and {Tuomi}, Mikko and {Teske}, Johanna and {Arriagada}, Pamela and {Diaz}, Matias and {Holden}, Brad and {Keiser}, Sandy},
        title = "{The LCES HIRES/Keck Precision Radial Velocity Exoplanet Survey}",
      journal = {\aj},
         year = 2017,
        month = may,
       volume = {153},
       number = {5},
          eid = {208},
        pages = {208},
          doi = {10.3847/1538-3881/aa66ca},
archivePrefix = {arXiv},
       eprint = {1702.03571},
 primaryClass = {astro-ph.EP},
       adsurl = {https://ui.adsabs.harvard.edu/abs/2017AJ....153..208B}
}

@ARTICLE{Teklu2025,
       author = {{Teklu}, Jerusalem T. and {Perdelwitz}, Volker and {Butler}, Robert Paul and {Trifonov}, Trifon and {Vogt}, Steven S. and {Mukhija}, Deepa and {Tal-Or}, Lev},
        title = "{An updated catalog of HIRES/Keck radial velocity measurements: Including Ca II H\&K measurements}",
      journal = {\aap},
         year = 2025,
        month = oct,
       volume = {702},
          eid = {A68},
        pages = {A68},
          doi = {10.1051/0004-6361/202555034},
archivePrefix = {arXiv},
       eprint = {2509.06072},
 primaryClass = {astro-ph.EP},
       adsurl = {https://ui.adsabs.harvard.edu/abs/2025A&A...702A..68T}
}

@ARTICLE{Brandt2018,
	author = {{Brandt}, Timothy D.},
	title = "{The Hipparcos-Gaia Catalog of Accelerations}",
	journal = {\apjs},
	year = 2018,
	month = dec,
	volume = {239},
	number = {2},
	eid = {31},
	pages = {31},
	doi = {10.3847/1538-4365/aaec06},
	archivePrefix = {arXiv},
	eprint = {1811.07283},
	primaryClass = {astro-ph.SR},
	adsurl = {https://ui.adsabs.harvard.edu/abs/2018ApJS..239...31B}
}

@ARTICLE{Brandt2021,
	author = {{Brandt}, Timothy D.},
	title = "{The Hipparcos-Gaia Catalog of Accelerations: Gaia EDR3 Edition}",
	journal = {\apjs},
	year = 2021,
	month = jun,
	volume = {254},
	number = {2},
	eid = {42},
	pages = {42},
	doi = {10.3847/1538-4365/abf93c},
	archivePrefix = {arXiv},
	eprint = {2105.11662},
	primaryClass = {astro-ph.GA},
	adsurl = {https://ui.adsabs.harvard.edu/abs/2021ApJS..254...42B}
}

@ARTICLE{Hipparcos,
       author = {{Perryman}, M.~A.~C. and {Lindegren}, L. and {Kovalevsky}, J. and {Hoeg}, E. and {Bastian}, U. and {Bernacca}, P.~L. and {Cr{\'e}z{\'e}}, M. and {Donati}, F. and {Grenon}, M. and {Grewing}, M. and {van Leeuwen}, F. and {van der Marel}, H. and {Mignard}, F. and {Murray}, C.~A. and {Le Poole}, R.~S. and {Schrijver}, H. and {Turon}, C. and {Arenou}, F. and {Froeschl{\'e}}, M. and {Petersen}, C.~S.},
        title = "{The HIPPARCOS Catalogue}",
      journal = {\aap},
         year = 1997,
        month = jul,
       volume = {323},
        pages = {L49-L52},
       adsurl = {https://ui.adsabs.harvard.edu/abs/1997A&A...323L..49P}
}

@ARTICLE{Gaia,
	author = {{Gaia Collaboration} and others},
	title = "{The Gaia mission}",
	journal = {\aap},
	year = 2016,
	month = nov,
	volume = {595},
	eid = {A1},
	pages = {A1},
	doi = {10.1051/0004-6361/201629272},
	archivePrefix = {arXiv},
	eprint = {1609.04153},
	primaryClass = {astro-ph.IM},
	adsurl = {https://ui.adsabs.harvard.edu/abs/2016A&A...595A...1G}
}

@ARTICLE{GaiaDR3,
       author = {{Gaia Collaboration} and others},
        title = "{Gaia Data Release 3. Summary of the content and survey properties}",
      journal = {\aap},
         year = 2023,
        month = jun,
       volume = {674},
          eid = {A1},
        pages = {A1},
          doi = {10.1051/0004-6361/202243940},
archivePrefix = {arXiv},
       eprint = {2208.00211},
 primaryClass = {astro-ph.GA},
       adsurl = {https://ui.adsabs.harvard.edu/abs/2023A&A...674A...1G}
}

@ARTICLE{Tycho2,
       author = {{H{\o}g}, E. and {Fabricius}, C. and {Makarov}, V.~V. and {Urban}, S. and {Corbin}, T. and {Wycoff}, G. and {Bastian}, U. and {Schwekendiek}, P. and {Wicenec}, A.},
        title = "{The Tycho-2 catalogue of the 2.5 million brightest stars}",
      journal = {\aap},
         year = 2000,
        month = mar,
       volume = {355},
        pages = {L27-L30},
       adsurl = {https://ui.adsabs.harvard.edu/abs/2000A&A...355L..27H}
}

@ARTICLE{2MASS,
	author = {{Skrutskie}, M.~F. and {Cutri}, R.~M. and {Stiening}, R. and {Weinberg}, M.~D. and {Schneider}, S. and {Carpenter}, J.~M. and {Beichman}, C. and {Capps}, R. and {Chester}, T. and {Elias}, J. and {Huchra}, J. and {Liebert}, J. and {Lonsdale}, C. and {Monet}, D.~G. and {Price}, S. and {Seitzer}, P. and {Jarrett}, T. and {Kirkpatrick}, J.~D. and {Gizis}, J.~E. and {Howard}, E. and {Evans}, T. and {Fowler}, J. and {Fullmer}, L. and {Hurt}, R. and {Light}, R. and {Kopan}, E.~L. and {Marsh}, K.~A. and {McCallon}, H.~L. and {Tam}, R. and {Van Dyk}, S. and {Wheelock}, S.},
	title = "{The Two Micron All Sky Survey (2MASS)}",
	journal = {\aj},
	year = 2006,
	month = feb,
	volume = {131},
	number = {2},
	pages = {1163-1183},
	doi = {10.1086/498708},
	adsurl = {https://ui.adsabs.harvard.edu/abs/2006AJ....131.1163S}
}

@dataset{AllWISE,
       author = {{Wright}, Edward L. and {Eisenhardt}, Peter R.~M. and {Mainzer}, Amy K. and {Ressler}, Michael E. and {Cutri}, Roc M. and {Jarrett}, Thomas and {Kirkpatrick}, J. Davy and {Padgett}, Deborah and {McMillan}, Robert S. and {Skrutskie}, Michael and {Stanford}, S.~A. and {Cohen}, Martin and {Walker}, Russell G. and {Mather}, John C. and {Leisawitz}, David and {Gautier}, III, Thomas N. and {McLean}, Ian and {Benford}, Dominic and {Lonsdale}, Carol J. and {Blain}, Andrew and {Mendez}, Bryan and {Irace}, William R. and {Duval}, Valerie and {Liu}, Fengchuan and {Royer}, Don and {Heinrichsen}, Ingolf and {Howard}, Joan and {Shannon}, Mark and {Kendall}, Martha and {Walsh}, Amy L. and {Larsen}, Mark and {Cardon}, Joel G. and {Schick}, Scott and {Schwalm}, Mark and {Abid}, Mohamed and {Fabinsky}, Beth and {Naes}, Larry and {Tsai}, ChaoWei},
        title = "{AllWISE Source Catalog}",
 howpublished = {NASA IPAC DataSet, IRSA1},
         year = 2019,
        month = jan,
          doi = {10.26131/IRSA1},
       adsurl = {https://ui.adsabs.harvard.edu/abs/2019ipac.data...I1W}
}

@ARTICLE{Gray2003,
       author = {{Gray}, R.~O. and {Corbally}, C.~J. and {Garrison}, R.~F. and {McFadden}, M.~T. and {Robinson}, P.~E.},
        title = "{Contributions to the Nearby Stars (NStars) Project: Spectroscopy of Stars Earlier than M0 within 40 Parsecs: The Northern Sample. I.}",
      journal = {\aj},
         year = 2003,
        month = oct,
       volume = {126},
       number = {4},
        pages = {2048-2059},
          doi = {10.1086/378365},
archivePrefix = {arXiv},
       eprint = {astro-ph/0308182},
 primaryClass = {astro-ph},
       adsurl = {https://ui.adsabs.harvard.edu/abs/2003AJ....126.2048G}
}

@ARTICLE{Rice2020,
       author = {{Rice}, Malena and {Brewer}, John M.},
        title = "{Stellar Characterization of Keck HIRES Spectra with The Cannon}",
      journal = {\apj},
         year = 2020,
        month = aug,
       volume = {898},
       number = {2},
          eid = {119},
        pages = {119},
          doi = {10.3847/1538-4357/ab9f96},
archivePrefix = {arXiv},
       eprint = {2007.02942},
 primaryClass = {astro-ph.EP},
       adsurl = {https://ui.adsabs.harvard.edu/abs/2020ApJ...898..119R}
}

@ARTICLE{Soubiran2018,
       author = {{Soubiran}, C. and {Jasniewicz}, G. and {Chemin}, L. and {Zurbach}, C. and {Brouillet}, N. and {Panuzzo}, P. and {Sartoretti}, P. and {Katz}, D. and {Le Campion}, J.-F. and {Marchal}, O. and {Hestroffer}, D. and {Th{\'e}venin}, F. and {Crifo}, F. and {Udry}, S. and {Cropper}, M. and {Seabroke}, G. and {Viala}, Y. and {Benson}, K. and {Blomme}, R. and {Jean-Antoine}, A. and {Huckle}, H. and {Smith}, M. and {Baker}, S.~G. and {Damerdji}, Y. and {Dolding}, C. and {Fr{\'e}mat}, Y. and {Gosset}, E. and {Guerrier}, A. and {Guy}, L.~P. and {Haigron}, R. and {Jan{\ss}en}, K. and {Plum}, G. and {Fabre}, C. and {Lasne}, Y. and {Pailler}, F. and {Panem}, C. and {Riclet}, F. and {Royer}, F. and {Tauran}, G. and {Zwitter}, T. and {Gueguen}, A. and {Turon}, C.},
        title = "{Gaia Data Release 2. The catalogue of radial velocity standard stars}",
      journal = {\aap},
         year = 2018,
        month = aug,
       volume = {616},
          eid = {A7},
        pages = {A7},
          doi = {10.1051/0004-6361/201832795},
archivePrefix = {arXiv},
       eprint = {1804.09370},
 primaryClass = {astro-ph.GA},
       adsurl = {https://ui.adsabs.harvard.edu/abs/2018A&A...616A...7S}
}

@ARTICLE{BailerJones2021,
       author = {{Bailer-Jones}, C.~A.~L. and {Rybizki}, J. and {Fouesneau}, M. and {Demleitner}, M. and {Andrae}, R.},
        title = "{Estimating Distances from Parallaxes. V. Geometric and Photogeometric Distances to 1.47 Billion Stars in Gaia Early Data Release 3}",
      journal = {\aj},
         year = 2021,
        month = mar,
       volume = {161},
       number = {3},
          eid = {147},
        pages = {147},
          doi = {10.3847/1538-3881/abd806},
archivePrefix = {arXiv},
       eprint = {2012.05220},
 primaryClass = {astro-ph.SR},
       adsurl = {https://ui.adsabs.harvard.edu/abs/2021AJ....161..147B}
}

@ARTICLE{Valenti2005,
       author = {{Valenti}, Jeff A. and {Fischer}, Debra A.},
        title = "{Spectroscopic Properties of Cool Stars (SPOCS). I. 1040 F, G, and K Dwarfs from Keck, Lick, and AAT Planet Search Programs}",
      journal = {\apjs},
         year = 2005,
        month = jul,
       volume = {159},
       number = {1},
        pages = {141-166},
          doi = {10.1086/430500},
       adsurl = {https://ui.adsabs.harvard.edu/abs/2005ApJS..159..141V}
}

@ARTICLE{Takeda2007,
       author = {{Takeda}, Genya and {Ford}, Eric B. and {Sills}, Alison and {Rasio}, Frederic A. and {Fischer}, Debra A. and {Valenti}, Jeff A.},
        title = "{Structure and Evolution of Nearby Stars with Planets. II. Physical Properties of \raisebox{-0.5ex}\textasciitilde1000 Cool Stars from the SPOCS Catalog}",
      journal = {\apjs},
         year = 2007,
        month = feb,
       volume = {168},
       number = {2},
        pages = {297-318},
          doi = {10.1086/509763},
archivePrefix = {arXiv},
       eprint = {astro-ph/0607235},
 primaryClass = {astro-ph},
       adsurl = {https://ui.adsabs.harvard.edu/abs/2007ApJS..168..297T}
}

@ARTICLE{daSilva2015,
       author = {{\noop{Silva}}{da Silva}, Ronaldo and {Milone}, Andr{\'e} de C. and {Rocha-Pinto}, Helio J.},
        title = "{Homogeneous abundance analysis of FGK dwarf, subgiant, and giant stars with and without giant planets}",
      journal = {\aap},
         year = 2015,
        month = aug,
       volume = {580},
          eid = {A24},
        pages = {A24},
          doi = {10.1051/0004-6361/201525770},
       adsurl = {https://ui.adsabs.harvard.edu/abs/2015A&A...580A..24D}
}

@ARTICLE{Johnson1987,
       author = {{Johnson}, Dean R.~H. and {Soderblom}, David R.},
        title = "{Calculating Galactic Space Velocities and Their Uncertainties, with an Application to the Ursa Major Group}",
      journal = {\aj},
         year = 1987,
        month = apr,
       volume = {93},
        pages = {864},
          doi = {10.1086/114370},
       adsurl = {https://ui.adsabs.harvard.edu/abs/1987AJ.....93..864J}
}

@ARTICLE{Schonrich2010,
       author = {{Sch{\"o}nrich}, Ralph and {Binney}, James and {Dehnen}, Walter},
        title = "{Local kinematics and the local standard of rest}",
      journal = {\mnras},
         year = 2010,
        month = apr,
       volume = {403},
       number = {4},
        pages = {1829-1833},
          doi = {10.1111/j.1365-2966.2010.16253.x},
archivePrefix = {arXiv},
       eprint = {0912.3693},
 primaryClass = {astro-ph.GA},
       adsurl = {https://ui.adsabs.harvard.edu/abs/2010MNRAS.403.1829S}
}

@ARTICLE{Bensby2003,
       author = {{Bensby}, T. and {Feltzing}, S. and {Lundstr{\"o}m}, I.},
        title = "{Elemental abundance trends in the Galactic thin and thick disks as traced by nearby F and G dwarf stars}",
      journal = {\aap},
         year = 2003,
        month = nov,
       volume = {410},
        pages = {527-551},
          doi = {10.1051/0004-6361:20031213},
       adsurl = {https://ui.adsabs.harvard.edu/abs/2003A&A...410..527B}
}

@ARTICLE{Hayden2017,
       author = {{Hayden}, M.~R. and {Recio-Blanco}, A. and {de Laverny}, P. and {Mikolaitis}, S. and {Worley}, C.~C.},
        title = "{The AMBRE project: The thick thin disk and thin thick disk of the Milky Way}",
      journal = {\aap},
         year = 2017,
        month = dec,
       volume = {608},
          eid = {L1},
        pages = {L1},
          doi = {10.1051/0004-6361/201731494},
archivePrefix = {arXiv},
       eprint = {1712.02358},
 primaryClass = {astro-ph.GA},
       adsurl = {https://ui.adsabs.harvard.edu/abs/2017A&A...608L...1H}
}

@ARTICLE{AlmeidaFernandes2018,
       author = {{Almeida-Fernandes}, F. and {Rocha-Pinto}, H.~J.},
        title = "{A method to estimate stellar ages from kinematical data}",
      journal = {\mnras},
         year = 2018,
        month = may,
       volume = {476},
       number = {1},
        pages = {184-197},
          doi = {10.1093/mnras/sty119},
archivePrefix = {arXiv},
       eprint = {1801.04046},
 primaryClass = {astro-ph.SR},
       adsurl = {https://ui.adsabs.harvard.edu/abs/2018MNRAS.476..184A}
}

@ARTICLE{Veyette2018,
       author = {{Veyette}, Mark J. and {Muirhead}, Philip S.},
        title = "{Chemo-kinematic Ages of Eccentric-planet-hosting M Dwarf Stars}",
      journal = {\apj},
         year = 2018,
        month = aug,
       volume = {863},
       number = {2},
          eid = {166},
        pages = {166},
          doi = {10.3847/1538-4357/aad40e},
archivePrefix = {arXiv},
       eprint = {1807.06017},
 primaryClass = {astro-ph.SR},
       adsurl = {https://ui.adsabs.harvard.edu/abs/2018ApJ...863..166V}
}

@ARTICLE{Dotter2016,
       author = {{Dotter}, Aaron},
        title = "{MESA Isochrones and Stellar Tracks (MIST) 0: Methods for the Construction of Stellar Isochrones}",
      journal = {\apjs},
         year = 2016,
        month = jan,
       volume = {222},
       number = {1},
          eid = {8},
        pages = {8},
          doi = {10.3847/0067-0049/222/1/8},
archivePrefix = {arXiv},
       eprint = {1601.05144},
 primaryClass = {astro-ph.SR},
       adsurl = {https://ui.adsabs.harvard.edu/abs/2016ApJS..222....8D}
}

@ARTICLE{Choi2016,
       author = {{Choi}, Jieun and {Dotter}, Aaron and {Conroy}, Charlie and {Cantiello}, Matteo and {Paxton}, Bill and {Johnson}, Benjamin D.},
        title = "{Mesa Isochrones and Stellar Tracks (MIST). I. Solar-scaled Models}",
      journal = {\apj},
         year = 2016,
        month = jun,
       volume = {823},
       number = {2},
          eid = {102},
        pages = {102},
          doi = {10.3847/0004-637X/823/2/102},
archivePrefix = {arXiv},
       eprint = {1604.08592},
 primaryClass = {astro-ph.SR},
       adsurl = {https://ui.adsabs.harvard.edu/abs/2016ApJ...823..102C}
}

@ARTICLE{Venner2026,
       author = {{Venner}, Alexander and {Vanderburg}, Andrew and {Huang}, Chelsea X. and {Dholakia}, Shishir and {Schwengeler}, Hans Martin and {Howell}, Steve B. and {Wittenmyer}, Robert A. and {Kristiansen}, Martti H. and {Omohundro}, Mark and {Terentev}, Ivan A.},
        title = "{A Cool Earth-sized Planet Candidate Transiting a Tenth Magnitude K-dwarf From K2}",
      journal = {\apjl},
         year = 2026,
        month = feb,
       volume = {997},
       number = {2},
          eid = {L38},
        pages = {L38},
          doi = {10.3847/2041-8213/adf06f},
archivePrefix = {arXiv},
       eprint = {2601.19870},
 primaryClass = {astro-ph.EP},
       adsurl = {https://ui.adsabs.harvard.edu/abs/2026ApJ...997L..38V}
}

@ARTICLE{Bidelman1951,
       author = {{Bidelman}, William P. and {Keenan}, Philip C.},
        title = "{The Ba II Stars.}",
      journal = {\apj},
         year = 1951,
        month = nov,
       volume = {114},
        pages = {473},
          doi = {10.1086/145488},
       adsurl = {https://ui.adsabs.harvard.edu/abs/1951ApJ...114..473B}
}

@ARTICLE{Keenan1954,
       author = {{Keenan}, Philip C.},
        title = "{Classification of the S-Type Stars.}",
      journal = {\apj},
         year = 1954,
        month = nov,
       volume = {120},
        pages = {484},
          doi = {10.1086/145937},
       adsurl = {https://ui.adsabs.harvard.edu/abs/1954ApJ...120..484K}
}

@ARTICLE{Jorissen2019,
       author = {{Jorissen}, A. and {Boffin}, H.~M.~J. and {Karinkuzhi}, D. and {Van Eck}, S. and {Escorza}, A. and {Shetye}, S. and {Van Winckel}, H.},
        title = "{Barium and related stars, and their white-dwarf companions. I. Giant stars}",
      journal = {\aap},
         year = 2019,
        month = jun,
       volume = {626},
          eid = {A127},
        pages = {A127},
          doi = {10.1051/0004-6361/201834630},
archivePrefix = {arXiv},
       eprint = {1904.03975},
 primaryClass = {astro-ph.SR},
       adsurl = {https://ui.adsabs.harvard.edu/abs/2019A&A...626A.127J}
}

@ARTICLE{Escorza2019,
       author = {{Escorza}, A. and {Karinkuzhi}, D. and {Jorissen}, A. and {Siess}, L. and {Van Winckel}, H. and {Pourbaix}, D. and {Johnston}, C. and {Miszalski}, B. and {Oomen}, G.-M. and {Abdul-Masih}, M. and {Boffin}, H.~M.~J. and {North}, P. and {Manick}, R. and {Shetye}, S. and {Miko{\l}ajewska}, J.},
        title = "{Barium and related stars, and their white-dwarf companions. II. Main-sequence and subgiant starss}",
      journal = {\aap},
         year = 2019,
        month = jun,
       volume = {626},
          eid = {A128},
        pages = {A128},
          doi = {10.1051/0004-6361/201935390},
archivePrefix = {arXiv},
       eprint = {1904.04095},
 primaryClass = {astro-ph.SR},
       adsurl = {https://ui.adsabs.harvard.edu/abs/2019A&A...626A.128E}
}

@ARTICLE{McClure1980,
       author = {{McClure}, R.~D. and {Fletcher}, J.~M. and {Nemec}, J.~M.},
        title = "{The binary nature of the barium stars.}",
      journal = {\apjl},
         year = 1980,
        month = may,
       volume = {238},
        pages = {L35-L38},
          doi = {10.1086/183252},
       adsurl = {https://ui.adsabs.harvard.edu/abs/1980ApJ...238L..35M}
}

@ARTICLE{McClure1984,
       author = {{McClure}, R.~D.},
        title = "{The barium stars.}",
      journal = {\pasp},
         year = 1984,
        month = feb,
       volume = {96},
        pages = {117-127},
          doi = {10.1086/131310},
       adsurl = {https://ui.adsabs.harvard.edu/abs/1984PASP...96..117M}
}

@ARTICLE{Kraus2016ImpactOfStellarMult,
   author = {{Kraus}, A.~L. and {Ireland}, M.~J. and {Huber}, D. and {Mann}, A.~W. and 
	{Dupuy}, T.~J.},
    title = "{The Impact of Stellar Multiplicity on Planetary Systems. I. The Ruinous Influence of Close Binary Companions}",
  journal = {\aj},
archivePrefix = "arXiv",
   eprint = {1604.05744},
 primaryClass = "astro-ph.EP",
     year = 2016,
    month = jul,
   volume = 152,
      eid = {8},
    pages = {8},
      doi = {10.3847/0004-6256/152/1/8},
   adsurl = {http://adsabs.harvard.edu/abs/2016AJ....152....8K}
}

@ARTICLE{Metchev2009YoungSolarAnalogs,
   author = {{Metchev}, S.~A. and {Hillenbrand}, L.~A.},
    title = "{The Palomar/Keck Adaptive Optics Survey of Young Solar Analogs: Evidence for a Universal Companion Mass Function}",
  journal = {\apjs},
archivePrefix = "arXiv",
   eprint = {0808.2982},
     year = 2009,
    month = mar,
   volume = 181,
    pages = {62-109},
      doi = {10.1088/0067-0049/181/1/62},
   adsurl = {http://adsabs.harvard.edu/abs/2009ApJS..181...62M}
}

@ARTICLE{Wizinowich2000,
       author = {{Wizinowich}, P. and {Acton}, D.~S. and {Shelton}, C. and {Stomski}, P.
        and {Gathright}, J. and {Ho}, K. and {Lupton}, W. and {Tsubota},
        K. and {Lai}, O. and {Max}, C. and {Brase}, J. and {An}, J. and
        {Avicola}, K. and {Olivier}, S. and {Gavel}, D. and {Macintosh},
        B. and {Ghez}, A. and {Larkin}, J.},
        title = "{First Light Adaptive Optics Images from the Keck II Telescope: A New Era
        of High Angular Resolution Imagery}",
      journal = {Publications of the Astronomical Society of the Pacific},
         year = 2000,
        month = Mar,
       volume = {112},
        pages = {315-319},
          doi = {10.1086/316543},
       adsurl = {https://ui.adsabs.harvard.edu/#abs/2000PASP..112..315W}
}

@ARTICLE{Yelda2010NIRC2Distortion,
   author = {{Yelda}, S. and {Lu}, J.~R. and {Ghez}, A.~M. and {Clarkson}, W. and 
	{Anderson}, J. and {Do}, T. and {Matthews}, K.},
    title = "{Improving Galactic Center Astrometry by Reducing the Effects of Geometric Distortion}",
  journal = {\apj},
archivePrefix = "arXiv",
   eprint = {1010.0064},
     year = 2010,
    month = dec,
   volume = 725,
      eid = {331-352},
    pages = {331-352},
      doi = {10.1088/0004-637X/725/1/331},
   adsurl = {http://adsabs.harvard.edu/abs/2010ApJ...725..331Y}
}

@ARTICLE{Service2016NIRC2Distortion,
       author = {{Service}, M. and {Lu}, J.~R. and {Campbell}, R. and {Sitarski}, B.~N. and {Ghez}, A.~M. and {Anderson}, J.},
        title = "{A New Distortion Solution for NIRC2 on the Keck II Telescope}",
      journal = {\pasp},
         year = 2016,
        month = sep,
       volume = {128},
       number = {967},
        pages = {095004},
          doi = {10.1088/1538-3873/128/967/095004},
       adsurl = {https://ui.adsabs.harvard.edu/abs/2016PASP..128i5004S}
}

@ARTICLE{Pearce2019GSC6214-210,
       author = {{Pearce}, Logan A. and {Kraus}, Adam L. and {Dupuy}, Trent J. and {Ireland
        }, Michael J. and {Rizzuto}, Aaron C. and {Bowler}, Brendan P. and
         {Birchall}, Eloise K. and {Wallace}, Alexander L.},
        title = "{Orbital Motion of the Wide Planetary-mass Companion GSC 6214-210 b: No Evidence for Dynamical Scattering}",
      journal = {\aj},
         year = "2019",
        month = "Feb",
       volume = {157},
       number = {2},
          eid = {71},
        pages = {71},
          doi = {10.3847/1538-3881/aafacb},
       adsurl = {https://ui.adsabs.harvard.edu/abs/2019AJ....157...71P}
}

@ARTICLE{Pearce2021BoyajiansStar,
       author = {{Pearce}, Logan A. and {Kraus}, Adam L. and {Dupuy}, Trent J. and {Mann}, Andrew W. and {Huber}, Daniel},
        title = "{Boyajian's Star B: The Co-moving Companion to KIC 8462852 A}",
      journal = {\apj},
         year = 2021,
        month = mar,
       volume = {909},
       number = {2},
          eid = {216},
        pages = {216},
          doi = {10.3847/1538-4357/abdd33},
archivePrefix = {arXiv},
       eprint = {2101.06313},
 primaryClass = {astro-ph.SR},
       adsurl = {https://ui.adsabs.harvard.edu/abs/2021ApJ...909..216P}
}

@ARTICLE{Vousden2016,
       author = {{Vousden}, W.~D. and {Farr}, W.~M. and {Mandel}, I.},
        title = "{Dynamic temperature selection for parallel tempering in Markov chain Monte Carlo simulations}",
      journal = {\mnras},
         year = 2016,
        month = jan,
       volume = {455},
       number = {2},
        pages = {1919-1937},
          doi = {10.1093/mnras/stv2422},
archivePrefix = {arXiv},
       eprint = {1501.05823},
 primaryClass = {stat.CO}
}

@ARTICLE{ForemanMackey2013,
       author = {{Foreman-Mackey}, Daniel and {Hogg}, David W. and {Lang}, Dustin and {Goodman}, Jonathan},
        title = "{emcee: The MCMC Hammer}",
      journal = {\pasp},
         year = 2013,
        month = mar,
       volume = {125},
       number = {925},
        pages = {306},
          doi = {10.1086/670067},
archivePrefix = {arXiv},
       eprint = {1202.3665},
 primaryClass = {astro-ph.IM}
}

@ARTICLE{astropyI,
       author = {{Astropy Collaboration} and {Robitaille}, Thomas P. and {Tollerud}, Erik J. and {Greenfield}, Perry and {Droettboom}, Michael and {Bray}, Erik and {Aldcroft}, Tom and {Davis}, Matt and {Ginsburg}, Adam and {Price-Whelan}, Adrian M. and {Kerzendorf}, Wolfgang E. and {Conley}, Alexander and {Crighton}, Neil and {Barbary}, Kyle and {Muna}, Demitri and {Ferguson}, Henry and {Grollier}, Fr{\'e}d{\'e}ric and {Parikh}, Madhura M. and {Nair}, Prasanth H. and {Unther}, Hans M. and {Deil}, Christoph and {Woillez}, Julien and {Conseil}, Simon and {Kramer}, Roban and {Turner}, James E.~H. and {Singer}, Leo and {Fox}, Ryan and {Weaver}, Benjamin A. and {Zabalza}, Victor and {Edwards}, Zachary I. and {Azalee Bostroem}, K. and {Burke}, D.~J. and {Casey}, Andrew R. and {Crawford}, Steven M. and {Dencheva}, Nadia and {Ely}, Justin and {Jenness}, Tim and {Labrie}, Kathleen and {Lim}, Pey Lian and {Pierfederici}, Francesco and {Pontzen}, Andrew and {Ptak}, Andy and {Refsdal}, Brian and {Servillat}, Mathieu and {Streicher}, Ole},
        title = "{Astropy: A community Python package for astronomy}",
      journal = {\aap},
         year = 2013,
        month = oct,
       volume = {558},
          eid = {A33},
        pages = {A33},
          doi = {10.1051/0004-6361/201322068},
archivePrefix = {arXiv},
       eprint = {1307.6212},
 primaryClass = {astro-ph.IM},
       adsurl = {https://ui.adsabs.harvard.edu/abs/2013A&A...558A..33A}
}

@ARTICLE{astropyII,
       author = {{Astropy Collaboration} and {Price-Whelan}, A.~M. and {Sip{\H{o}}cz}, B.~M. and {G{\"u}nther}, H.~M. and {Lim}, P.~L. and {Crawford}, S.~M. and {Conseil}, S. and {Shupe}, D.~L. and {Craig}, M.~W. and {Dencheva}, N. and {Ginsburg}, A. and {VanderPlas}, J.~T. and {Bradley}, L.~D. and {P{\'e}rez-Su{\'a}rez}, D. and {de Val-Borro}, M. and {Aldcroft}, T.~L. and {Cruz}, K.~L. and {Robitaille}, T.~P. and {Tollerud}, E.~J. and {Ardelean}, C. and {Babej}, T. and {Bach}, Y.~P. and {Bachetti}, M. and {Bakanov}, A.~V. and {Bamford}, S.~P. and {Barentsen}, G. and {Barmby}, P. and {Baumbach}, A. and {Berry}, K.~L. and {Biscani}, F. and {Boquien}, M. and {Bostroem}, K.~A. and {Bouma}, L.~G. and {Brammer}, G.~B. and {Bray}, E.~M. and {Breytenbach}, H. and {Buddelmeijer}, H. and {Burke}, D.~J. and {Calderone}, G. and {Cano Rodr{\'\i}guez}, J.~L. and {Cara}, M. and {Cardoso}, J.~V.~M. and {Cheedella}, S. and {Copin}, Y. and {Corrales}, L. and {Crichton}, D. and {D'Avella}, D. and {Deil}, C. and {Depagne}, {\'E}. and {Dietrich}, J.~P. and {Donath}, A. and {Droettboom}, M. and {Earl}, N. and {Erben}, T. and {Fabbro}, S. and {Ferreira}, L.~A. and {Finethy}, T. and {Fox}, R.~T. and {Garrison}, L.~H. and {Gibbons}, S.~L.~J. and {Goldstein}, D.~A. and {Gommers}, R. and {Greco}, J.~P. and {Greenfield}, P. and {Groener}, A.~M. and {Grollier}, F. and {Hagen}, A. and {Hirst}, P. and {Homeier}, D. and {Horton}, A.~J. and {Hosseinzadeh}, G. and {Hu}, L. and {Hunkeler}, J.~S. and {Ivezi{\'c}}, {\v{Z}}. and {Jain}, A. and {Jenness}, T. and {Kanarek}, G. and {Kendrew}, S. and {Kern}, N.~S. and {Kerzendorf}, W.~E. and {Khvalko}, A. and {King}, J. and {Kirkby}, D. and {Kulkarni}, A.~M. and {Kumar}, A. and {Lee}, A. and {Lenz}, D. and {Littlefair}, S.~P. and {Ma}, Z. and {Macleod}, D.~M. and {Mastropietro}, M. and {McCully}, C. and {Montagnac}, S. and {Morris}, B.~M. and {Mueller}, M. and {Mumford}, S.~J. and {Muna}, D. and {Murphy}, N.~A. and {Nelson}, S. and {Nguyen}, G.~H. and {Ninan}, J.~P. and {N{\"o}the}, M. and {Ogaz}, S. and {Oh}, S. and {Parejko}, J.~K. and {Parley}, N. and {Pascual}, S. and {Patil}, R. and {Patil}, A.~A. and {Plunkett}, A.~L. and {Prochaska}, J.~X. and {Rastogi}, T. and {Reddy Janga}, V. and {Sabater}, J. and {Sakurikar}, P. and {Seifert}, M. and {Sherbert}, L.~E. and {Sherwood-Taylor}, H. and {Shih}, A.~Y. and {Sick}, J. and {Silbiger}, M.~T. and {Singanamalla}, S. and {Singer}, L.~P. and {Sladen}, P.~H. and {Sooley}, K.~A. and {Sornarajah}, S. and {Streicher}, O. and {Teuben}, P. and {Thomas}, S.~W. and {Tremblay}, G.~R. and {Turner}, J.~E.~H. and {Terr{\'o}n}, V. and {van Kerkwijk}, M.~H. and {de la Vega}, A. and {Watkins}, L.~L. and {Weaver}, B.~A. and {Whitmore}, J.~B. and {Woillez}, J. and {Zabalza}, V. and {Astropy Contributors}},
        title = "{The Astropy Project: Building an Open-science Project and Status of the v2.0 Core Package}",
      journal = {\aj},
         year = 2018,
        month = sep,
       volume = {156},
       number = {3},
          eid = {123},
        pages = {123},
          doi = {10.3847/1538-3881/aabc4f},
archivePrefix = {arXiv},
       eprint = {1801.02634},
 primaryClass = {astro-ph.IM},
       adsurl = {https://ui.adsabs.harvard.edu/abs/2018AJ....156..123A}
}

@ARTICLE{astropyIII,
       author = {{Astropy Collaboration} and {Price-Whelan}, Adrian M. and {Lim}, Pey Lian and {Earl}, Nicholas and {Starkman}, Nathaniel and {Bradley}, Larry and {Shupe}, David L. and {Patil}, Aarya A. and {Corrales}, Lia and {Brasseur}, C.~E. and {N{\"o}the}, Maximilian and {Donath}, Axel and {Tollerud}, Erik and {Morris}, Brett M. and {Ginsburg}, Adam and {Vaher}, Eero and {Weaver}, Benjamin A. and {Tocknell}, James and {Jamieson}, William and {van Kerkwijk}, Marten H. and {Robitaille}, Thomas P. and {Merry}, Bruce and {Bachetti}, Matteo and {G{\"u}nther}, H. Moritz and {Aldcroft}, Thomas L. and {Alvarado-Montes}, Jaime A. and {Archibald}, Anne M. and {B{\'o}di}, Attila and {Bapat}, Shreyas and {Barentsen}, Geert and {Baz{\'a}n}, Juanjo and {Biswas}, Manish and {Boquien}, M{\'e}d{\'e}ric and {Burke}, D.~J. and {Cara}, Daria and {Cara}, Mihai and {Conroy}, Kyle E. and {Conseil}, Simon and {Craig}, Matthew W. and {Cross}, Robert M. and {Cruz}, Kelle L. and {D'Eugenio}, Francesco and {Dencheva}, Nadia and {Devillepoix}, Hadrien A.~R. and {Dietrich}, J{\"o}rg P. and {Eigenbrot}, Arthur Davis and {Erben}, Thomas and {Ferreira}, Leonardo and {Foreman-Mackey}, Daniel and {Fox}, Ryan and {Freij}, Nabil and {Garg}, Suyog and {Geda}, Robel and {Glattly}, Lauren and {Gondhalekar}, Yash and {Gordon}, Karl D. and {Grant}, David and {Greenfield}, Perry and {Groener}, Austen M. and {Guest}, Steve and {Gurovich}, Sebastian and {Handberg}, Rasmus and {Hart}, Akeem and {Hatfield-Dodds}, Zac and {Homeier}, Derek and {Hosseinzadeh}, Griffin and {Jenness}, Tim and {Jones}, Craig K. and {Joseph}, Prajwel and {Kalmbach}, J. Bryce and {Karamehmetoglu}, Emir and {Ka{\l}uszy{\'n}ski}, Miko{\l}aj and {Kelley}, Michael S.~P. and {Kern}, Nicholas and {Kerzendorf}, Wolfgang E. and {Koch}, Eric W. and {Kulumani}, Shankar and {Lee}, Antony and {Ly}, Chun and {Ma}, Zhiyuan and {MacBride}, Conor and {Maljaars}, Jakob M. and {Muna}, Demitri and {Murphy}, N.~A. and {Norman}, Henrik and {O'Steen}, Richard and {Oman}, Kyle A. and {Pacifici}, Camilla and {Pascual}, Sergio and {Pascual-Granado}, J. and {Patil}, Rohit R. and {Perren}, Gabriel I. and {Pickering}, Timothy E. and {Rastogi}, Tanuj and {Roulston}, Benjamin R. and {Ryan}, Daniel F. and {Rykoff}, Eli S. and {Sabater}, Jose and {Sakurikar}, Parikshit and {Salgado}, Jes{\'u}s and {Sanghi}, Aniket and {Saunders}, Nicholas and {Savchenko}, Volodymyr and {Schwardt}, Ludwig and {Seifert-Eckert}, Michael and {Shih}, Albert Y. and {Jain}, Anany Shrey and {Shukla}, Gyanendra and {Sick}, Jonathan and {Simpson}, Chris and {Singanamalla}, Sudheesh and {Singer}, Leo P. and {Singhal}, Jaladh and {Sinha}, Manodeep and {Sip{\H{o}}cz}, Brigitta M. and {Spitler}, Lee R. and {Stansby}, David and {Streicher}, Ole and {{\v{S}}umak}, Jani and {Swinbank}, John D. and {Taranu}, Dan S. and {Tewary}, Nikita and {Tremblay}, Grant R. and {de Val-Borro}, Miguel and {Van Kooten}, Samuel J. and {Vasovi{\'c}}, Zlatan and {Verma}, Shresth and {de Miranda Cardoso}, Jos{\'e} Vin{\'\i}cius and {Williams}, Peter K.~G. and {Wilson}, Tom J. and {Winkel}, Benjamin and {Wood-Vasey}, W.~M. and {Xue}, Rui and {Yoachim}, Peter and {Zhang}, Chen and {Zonca}, Andrea and {Astropy Project Contributors}},
        title = "{The Astropy Project: Sustaining and Growing a Community-oriented Open-source Project and the Latest Major Release (v5.0) of the Core Package}",
      journal = {\apj},
         year = 2022,
        month = aug,
       volume = {935},
       number = {2},
          eid = {167},
        pages = {167},
          doi = {10.3847/1538-4357/ac7c74},
archivePrefix = {arXiv},
       eprint = {2206.14220},
 primaryClass = {astro-ph.IM},
       adsurl = {https://ui.adsabs.harvard.edu/abs/2022ApJ...935..167A}
}

@software{william_o_balmer_2025_Backtracks,
  author       = {William O Balmer and
                  Gilles Otten and
                  Tomas Stolker},
  title        = {backtracks: a python package to compare relative
                   astrometry with background helical motion
                  },
  month        = feb,
  year         = 2025,
  publisher    = {Zenodo},
  version      = {v0.6.0},
  doi          = {10.5281/zenodo.14838370},
  url          = {https://doi.org/10.5281/zenodo.14838370},
  swhid        = {swh:1:dir:bede5e0d41dd3356bf01b7e2ce5d46bb929c2d51
                   ;origin=https://doi.org/10.5281/zenodo.14838369;vi
                   sit=swh:1:snp:f25aa919b9ff86fdce28c44856a57c834603
                   a2ca;anchor=swh:1:rel:cdf3f9e961033a83b66c0a354c5c
                   2316d2c4650e;path=wbalmer-backtracks-25e59c9
                  },
}

@ARTICLE{Lindegren2021,
       author = {{Lindegren}, L. and {Klioner}, S.~A. and {Hern{\'a}ndez}, J. and {Bombrun}, A. and {Ramos-Lerate}, M. and {Steidelm{\"u}ller}, H. and {Bastian}, U. and {Biermann}, M. and {de Torres}, A. and {Gerlach}, E. and {Geyer}, R. and {Hilger}, T. and {Hobbs}, D. and {Lammers}, U. and {McMillan}, P.~J. and {Stephenson}, C.~A. and {Casta{\~n}eda}, J. and {Davidson}, M. and {Fabricius}, C. and {Gracia-Abril}, G. and {Portell}, J. and {Rowell}, N. and {Teyssier}, D. and {Torra}, F. and {Bartolom{\'e}}, S. and {Clotet}, M. and {Garralda}, N. and {Gonz{\'a}lez-Vidal}, J.~J. and {Torra}, J. and {Abbas}, U. and {Altmann}, M. and {Anglada Varela}, E. and {Balaguer-N{\'u}{\~n}ez}, L. and {Balog}, Z. and {Barache}, C. and {Becciani}, U. and {Bernet}, M. and {Bertone}, S. and {Bianchi}, L. and {Bouquillon}, S. and {Brown}, A.~G.~A. and {Bucciarelli}, B. and {Busonero}, D. and {Butkevich}, A.~G. and {Buzzi}, R. and {Cancelliere}, R. and {Carlucci}, T. and {Charlot}, P. and {Cioni}, M.-R.~L. and {Crosta}, M. and {Crowley}, C. and {del Peloso}, E.~F. and {del Pozo}, E. and {Drimmel}, R. and {Esquej}, P. and {Fienga}, A. and {Fraile}, E. and {Gai}, M. and {Garcia-Reinaldos}, M. and {Guerra}, R. and {Hambly}, N.~C. and {Hauser}, M. and {Jan{\ss}en}, K. and {Jordan}, S. and {Kostrzewa-Rutkowska}, Z. and {Lattanzi}, M.~G. and {Liao}, S. and {Licata}, E. and {Lister}, T.~A. and {L{\"o}ffler}, W. and {Marchant}, J.~M. and {Masip}, A. and {Mignard}, F. and {Mints}, A. and {Molina}, D. and {Mora}, A. and {Morbidelli}, R. and {Murphy}, C.~P. and {Pagani}, C. and {Panuzzo}, P. and {Pe{\~n}alosa Esteller}, X. and {Poggio}, E. and {Re Fiorentin}, P. and {Riva}, A. and {Sagrist{\`a} Sell{\'e}s}, A. and {Sanchez Gimenez}, V. and {Sarasso}, M. and {Sciacca}, E. and {Siddiqui}, H.~I. and {Smart}, R.~L. and {Souami}, D. and {Spagna}, A. and {Steele}, I.~A. and {Taris}, F. and {Utrilla}, E. and {van Reeven}, W. and {Vecchiato}, A.},
        title = "{Gaia Early Data Release 3. The astrometric solution}",
      journal = {\aap},
         year = 2021,
        month = may,
       volume = {649},
          eid = {A2},
        pages = {A2},
          doi = {10.1051/0004-6361/202039709},
archivePrefix = {arXiv},
       eprint = {2012.03380},
 primaryClass = {astro-ph.IM},
       adsurl = {https://ui.adsabs.harvard.edu/abs/2021A&A...649A...2L}
}

@ARTICLE{Tremblay2016,
       author = {{Tremblay}, P.-E. and {Cummings}, J. and {Kalirai}, J.~S. and {G{\"a}nsicke}, B.~T. and {Gentile-Fusillo}, N. and {Raddi}, R.},
        title = "{The field white dwarf mass distribution}",
      journal = {\mnras},
         year = 2016,
        month = sep,
       volume = {461},
       number = {2},
        pages = {2100-2114},
          doi = {10.1093/mnras/stw1447},
archivePrefix = {arXiv},
       eprint = {1606.05292},
 primaryClass = {astro-ph.SR},
       adsurl = {https://ui.adsabs.harvard.edu/abs/2016MNRAS.461.2100T}
}

@ARTICLE{GCNS,
       author = {{Gaia Collaboration} and others},
        title = "{Gaia Early Data Release 3. The Gaia Catalogue of Nearby Stars}",
      journal = {\aap},
         year = 2021,
        month = may,
       volume = {649},
          eid = {A6},
        pages = {A6},
          doi = {10.1051/0004-6361/202039498},
archivePrefix = {arXiv},
       eprint = {2012.02061},
 primaryClass = {astro-ph.SR},
       adsurl = {https://ui.adsabs.harvard.edu/abs/2021A&A...649A...6G}
}

@ARTICLE{OBrien2024,
       author = {{O'Brien}, Mairi W. and {Tremblay}, P.-E. and {Klein}, B.~L. and {Koester}, D. and {Melis}, C. and {B{\'e}dard}, A. and {Cukanovaite}, E. and {Cunningham}, T. and {Doyle}, A.~E. and {G{\"a}nsicke}, B.~T. and {Gentile Fusillo}, N.~P. and {Hollands}, M.~A. and {McCleery}, J. and {Pelisoli}, I. and {Toonen}, S. and {Weinberger}, A.~J. and {Zuckerman}, B.},
        title = "{The 40 pc sample of white dwarfs from Gaia}",
      journal = {\mnras},
         year = 2024,
        month = jan,
       volume = {527},
       number = {3},
        pages = {8687-8705},
          doi = {10.1093/mnras/stad3773},
archivePrefix = {arXiv},
       eprint = {2312.02735},
 primaryClass = {astro-ph.SR},
       adsurl = {https://ui.adsabs.harvard.edu/abs/2024MNRAS.527.8687O}
}

@ARTICLE{orvara,
       author = {{Brandt}, Timothy D. and {Dupuy}, Trent J. and {Li}, Yiting and {Brandt}, G. Mirek and {Zeng}, Yunlin and {Michalik}, Daniel and {Bardalez Gagliuffi}, Daniella C. and {Raposo-Pulido}, Virginia},
        title = "{orvara: An Efficient Code to Fit Orbits Using Radial Velocity, Absolute, and/or Relative Astrometry}",
      journal = {\aj},
         year = 2021,
        month = nov,
       volume = {162},
       number = {5},
          eid = {186},
        pages = {186},
          doi = {10.3847/1538-3881/ac042e},
archivePrefix = {arXiv},
       eprint = {2105.11671},
 primaryClass = {astro-ph.IM},
       adsurl = {https://ui.adsabs.harvard.edu/abs/2021AJ....162..186B}
}

@ARTICLE{htof,
       author = {{Brandt}, G. Mirek and {Michalik}, Daniel and {Brandt}, Timothy D. and {Li}, Yiting and {Dupuy}, Trent J. and {Zeng}, Yunlin},
        title = "{htof: A New Open-source Tool for Analyzing Hipparcos, Gaia, and Future Astrometric Missions}",
      journal = {\aj},
         year = 2021,
        month = dec,
       volume = {162},
       number = {6},
          eid = {230},
        pages = {230},
          doi = {10.3847/1538-3881/ac12d0},
archivePrefix = {arXiv},
       eprint = {2109.06761},
 primaryClass = {astro-ph.IM},
       adsurl = {https://ui.adsabs.harvard.edu/abs/2021AJ....162..230B}
}

@ARTICLE{An2025,
       author = {{An}, Qier and {Brandt}, Timothy D. and {Brandt}, G. Mirek and {Venner}, Alexander},
        title = "{Orbits and Masses for 156 Companions from Combined Astrometry and Radial Velocities, and a Validation of Gaia Non-single-star Solutions}",
      journal = {\apjs},
         year = 2025,
        month = oct,
       volume = {280},
       number = {2},
          eid = {61},
        pages = {61},
          doi = {10.3847/1538-4365/adfa99},
archivePrefix = {arXiv},
       eprint = {2508.08374},
 primaryClass = {astro-ph.EP},
       adsurl = {https://ui.adsabs.harvard.edu/abs/2025ApJS..280...61A}
}

@ARTICLE{GaiaDR2,
       author = {{Gaia Collaboration} and others},
        title = "{Gaia Data Release 2. Summary of the contents and survey properties}",
      journal = {\aap},
         year = 2018,
        month = aug,
       volume = {616},
          eid = {A1},
        pages = {A1},
          doi = {10.1051/0004-6361/201833051},
archivePrefix = {arXiv},
       eprint = {1804.09365},
 primaryClass = {astro-ph.GA},
       adsurl = {https://ui.adsabs.harvard.edu/abs/2018A&A...616A...1G}
}

@ARTICLE{Li2021,
       author = {{Li}, Yiting and {Brandt}, Timothy D. and {Brandt}, G. Mirek and {Dupuy}, Trent J. and {Michalik}, Daniel and {Jensen-Clem}, Rebecca and {Zeng}, Yunlin and {Faherty}, Jacqueline and {Mitra}, Elena L.},
        title = "{Precise Masses and Orbits for Nine Radial-velocity Exoplanets}",
      journal = {\aj},
         year = 2021,
        month = dec,
       volume = {162},
       number = {6},
          eid = {266},
        pages = {266},
          doi = {10.3847/1538-3881/ac27ab},
archivePrefix = {arXiv},
       eprint = {2109.10422},
 primaryClass = {astro-ph.EP},
       adsurl = {https://ui.adsabs.harvard.edu/abs/2021AJ....162..266L}
}

@ARTICLE{OBrien2026,
       author = {{O'Brien}, Mairi W. and {Wilson}, David J. and {Tremblay}, Pier-Emmanuel and {G{\"a}nsicke}, Boris T. and {Byrne}, Conor M. and {Lagos-Vilches}, Felipe and {Pineda}, J. Sebastian and {Barraza-Jorquera}, Joaqu{\'\i}n A.},
        title = "{Direct detections of white dwarfs in four WD+dM post-common envelope binaries within 20 pc}",
      journal = {\mnras},
         year = 2026,
        month = aug,
       volume = {550},
       number = {2},
          eid = {stag1195},
        pages = {stag1195},
          doi = {10.1093/mnras/stag1195},
archivePrefix = {arXiv},
       eprint = {2607.11320},
 primaryClass = {astro-ph.SR},
       adsurl = {https://ui.adsabs.harvard.edu/abs/2026MNRAS.550g1195O}
}

@ARTICLE{Marigo2020,
       author = {{Marigo}, Paola and {Cummings}, Jeffrey D. and {Curtis}, Jason Lee and {Kalirai}, Jason and {Chen}, Yang and {Tremblay}, Pier-Emmanuel and {Ramirez-Ruiz}, Enrico and {Bergeron}, Pierre and {Bladh}, Sara and {Bressan}, Alessandro and {Girardi}, L{\'e}o and {Pastorelli}, Giada and {Trabucchi}, Michele and {Cheng}, Sihao and {Aringer}, Bernhard and {Tio}, Piero Dal},
        title = "{Carbon star formation as seen through the non-monotonic initial-final mass relation}",
      journal = {Nature Astronomy},
         year = 2020,
        month = jul,
       volume = {4},
        pages = {1102-1110},
          doi = {10.1038/s41550-020-1132-1},
archivePrefix = {arXiv},
       eprint = {2007.04163},
 primaryClass = {astro-ph.SR},
       adsurl = {https://ui.adsabs.harvard.edu/abs/2020NatAs...4.1102M}
}

@ARTICLE{corner,
       author = {{Foreman-Mackey}, Daniel},
        title = "{corner.py: Scatterplot matrices in Python}",
      journal = {The Journal of Open Source Software},
         year = 2016,
        month = jun,
       volume = {1},
        pages = {24},
          doi = {10.21105/joss.00024},
       adsurl = {https://ui.adsabs.harvard.edu/abs/2016JOSS....1...24F}
}

@ARTICLE{numpy,
       author = {{Harris}, Charles R. and {Millman}, K. Jarrod and {van der Walt}, St{\'e}fan J. and {Gommers}, Ralf and {Virtanen}, Pauli and {Cournapeau}, David and {Wieser}, Eric and {Taylor}, Julian and {Berg}, Sebastian and {Smith}, Nathaniel J. and {Kern}, Robert and {Picus}, Matti and {Hoyer}, Stephan and {van Kerkwijk}, Marten H. and {Brett}, Matthew and {Haldane}, Allan and {del R{\'\i}o}, Jaime Fern{\'a}ndez and {Wiebe}, Mark and {Peterson}, Pearu and {G{\'e}rard-Marchant}, Pierre and {Sheppard}, Kevin and {Reddy}, Tyler and {Weckesser}, Warren and {Abbasi}, Hameer and {Gohlke}, Christoph and {Oliphant}, Travis E.},
        title = "{Array programming with NumPy}",
      journal = {\nat},
         year = 2020,
        month = sep,
       volume = {585},
       number = {7825},
        pages = {357-362},
          doi = {10.1038/s41586-020-2649-2},
archivePrefix = {arXiv},
       eprint = {2006.10256},
 primaryClass = {cs.MS},
       adsurl = {https://ui.adsabs.harvard.edu/abs/2020Natur.585..357H}
}

@article{scipy,
  title={SciPy 1.0: fundamental algorithms for scientific computing in Python},
  author={Virtanen, Pauli and Gommers, Ralf and Oliphant, Travis E and Haberland, Matt and Reddy, Tyler and Cournapeau, David and Burovski, Evgeni and Peterson, Pearu and Weckesser, Warren and Bright, Jonathan and others},
  journal={Nature methods},
  volume={17},
  number={3},
  pages={261--272},
  year={2020},
  publisher={Nature Publishing Group US New York}
}

@ARTICLE{matplotlib,
       author = {{Hunter}, John D.},
        title = "{Matplotlib: A 2D Graphics Environment}",
      journal = {Computing in Science and Engineering},
         year = 2007,
        month = jan,
       volume = {9},
       number = {3},
        pages = {90-95},
          doi = {10.1109/MCSE.2007.55},
       adsurl = {https://ui.adsabs.harvard.edu/abs/2007CSE.....9...90H}
}

@ARTICLE{Cheng2025,
       author = {{Cheng}, Ho Wan and {Trifonov}, Trifon and {Lee}, Man Hoi and {Cantalloube}, Faustine and {Reffert}, Sabine and {Ramm}, David and {Quirrenbach}, Andreas},
        title = "{A retrograde planet in a tight binary star system with a white dwarf}",
      journal = {\nat},
         year = 2025,
        month = may,
       volume = {641},
       number = {8064},
        pages = {866-870},
          doi = {10.1038/s41586-025-09006-x},
       adsurl = {https://ui.adsabs.harvard.edu/abs/2025Natur.641..866C}
}

@ARTICLE{Gies2008,
       author = {{Gies}, D.~R. and {Dieterich}, S. and {Richardson}, N.~D. and {Riedel}, A.~R. and {B.~L. Team} and {McAlister}, H.~A. and {Bagnuolo}, Jr., W.~G. and {Grundstrom}, E.~D. and {{\v{S}}tefl}, S. and {Rivinius}, Th. and {Baade}, D.},
        title = "{A Spectroscopic Orbit for Regulus}",
      journal = {\apjl},
         year = 2008,
        month = aug,
       volume = {682},
       number = {2},
        pages = {L117},
          doi = {10.1086/591148},
archivePrefix = {arXiv},
       eprint = {0806.3473},
 primaryClass = {astro-ph},
       adsurl = {https://ui.adsabs.harvard.edu/abs/2008ApJ...682L.117G}
}

@ARTICLE{Rappaport2009,
       author = {{Rappaport}, S. and {Podsiadlowski}, Ph. and {Horev}, I.},
        title = "{The Past and Future History of Regulus}",
      journal = {\apj},
         year = 2009,
        month = jun,
       volume = {698},
       number = {1},
        pages = {666-675},
          doi = {10.1088/0004-637X/698/1/666},
archivePrefix = {arXiv},
       eprint = {0904.0395},
 primaryClass = {astro-ph.SR},
       adsurl = {https://ui.adsabs.harvard.edu/abs/2009ApJ...698..666R}
}

@ARTICLE{Gies2020,
       author = {{Gies}, Douglas R. and {Lester}, Kathryn V. and {Wang}, Luqian and {Couperus}, Andrew and {Shepard}, Katherine and {Neiner}, Coralie and {Wade}, Gregg A. and {Dunham}, David W. and {Dunham}, Joan B.},
        title = "{Spectroscopic Detection of the Pre-White Dwarf Companion of Regulus}",
      journal = {\apj},
         year = 2020,
        month = oct,
       volume = {902},
       number = {1},
          eid = {25},
        pages = {25},
          doi = {10.3847/1538-4357/abb372},
archivePrefix = {arXiv},
       eprint = {2009.02409},
 primaryClass = {astro-ph.SR},
       adsurl = {https://ui.adsabs.harvard.edu/abs/2020ApJ...902...25G}
}

@ARTICLE{WDS,
       author = {{Mason}, Brian D. and {Wycoff}, Gary L. and {Hartkopf}, William I. and {Douglass}, Geoffrey G. and {Worley}, Charles E.},
        title = "{The 2001 US Naval Observatory Double Star CD-ROM. I. The Washington Double Star Catalog}",
      journal = {\aj},
         year = 2001,
        month = dec,
       volume = {122},
       number = {6},
        pages = {3466-3471},
          doi = {10.1086/323920},
       adsurl = {https://ui.adsabs.harvard.edu/abs/2001AJ....122.3466M}
}

@ARTICLE{Bond2017,
       author = {{Bond}, Howard E. and {Schaefer}, Gail H. and {Gilliland}, Ronald L. and {Holberg}, Jay B. and {Mason}, Brian D. and {Lindenblad}, Irving W. and {Seitz-McLeese}, Miranda and {Arnett}, W. David and {Demarque}, Pierre and {Spada}, Federico and {Young}, Patrick A. and {Barstow}, Martin A. and {Burleigh}, Matthew R. and {Gudehus}, Donald},
        title = "{The Sirius System and Its Astrophysical Puzzles: Hubble Space Telescope and Ground-based Astrometry}",
      journal = {\apj},
         year = 2017,
        month = may,
       volume = {840},
       number = {2},
          eid = {70},
        pages = {70},
          doi = {10.3847/1538-4357/aa6af8},
archivePrefix = {arXiv},
       eprint = {1703.10625},
 primaryClass = {astro-ph.SR},
       adsurl = {https://ui.adsabs.harvard.edu/abs/2017ApJ...840...70B}
}

@ARTICLE{Provencal2002,
       author = {{Provencal}, J.~L. and {Shipman}, H.~L. and {Koester}, Detlev and {Wesemael}, F. and {Bergeron}, P.},
        title = "{Procyon B: Outside the Iron Box}",
      journal = {\apj},
         year = 2002,
        month = mar,
       volume = {568},
       number = {1},
        pages = {324-334},
          doi = {10.1086/338769},
       adsurl = {https://ui.adsabs.harvard.edu/abs/2002ApJ...568..324P}
}

@ARTICLE{Bond2015,
       author = {{Bond}, Howard E. and {Gilliland}, Ronald L. and {Schaefer}, Gail H. and {Demarque}, Pierre and {Girard}, Terrence M. and {Holberg}, Jay B. and {Gudehus}, Donald and {Mason}, Brian D. and {Kozhurina-Platais}, Vera and {Burleigh}, Matthew R. and {Barstow}, Martin A. and {Nelan}, Edmund P.},
        title = "{Hubble Space Telescope Astrometry of the Procyon System}",
      journal = {\apj},
         year = 2015,
        month = nov,
       volume = {813},
       number = {2},
          eid = {106},
        pages = {106},
          doi = {10.1088/0004-637X/813/2/106},
archivePrefix = {arXiv},
       eprint = {1510.00485},
 primaryClass = {astro-ph.SR},
       adsurl = {https://ui.adsabs.harvard.edu/abs/2015ApJ...813..106B}
}

@ARTICLE{Bond2017.40Eri,
       author = {{Bond}, Howard E. and {Bergeron}, P. and {B{\'e}dard}, A.},
        title = "{Astrophysical Implications of a New Dynamical Mass for the Nearby White Dwarf 40 Eridani B}",
      journal = {\apj},
         year = 2017,
        month = oct,
       volume = {848},
       number = {1},
          eid = {16},
        pages = {16},
          doi = {10.3847/1538-4357/aa8a63},
archivePrefix = {arXiv},
       eprint = {1709.00478},
 primaryClass = {astro-ph.SR},
       adsurl = {https://ui.adsabs.harvard.edu/abs/2017ApJ...848...16B}
}

@ARTICLE{vanBiesbroeck1961,
	author = {{\noop{Biesbroeck}}{van Biesbroeck}, G.},
	title = "{A search for Stars of Low Luminosity .}",
	journal = {\aj},
	year = 1961,
	month = nov,
	volume = {66},
	pages = {528-530},
	doi = {10.1086/108457},
	adsurl = {https://ui.adsabs.harvard.edu/abs/1961AJ.....66..528V}
}

@ARTICLE{Hawley1996,
	author = {{Hawley}, Suzanne L. and {Gizis}, John E. and {Reid}, I. Neill},
	title = "{The Palomar/MSU Nearby Star Spectroscopic Survey.II.The Southern M Dwarfs and Investigation of Magnetic Activity}",
	journal = {\aj},
	year = 1996,
	month = dec,
	volume = {112},
	pages = {2799},
	doi = {10.1086/118222},
	adsurl = {https://ui.adsabs.harvard.edu/abs/1996AJ....112.2799H}
}

@ARTICLE{Henry2002,
	author = {{Henry}, Todd J. and {Walkowicz}, Lucianne M. and {Barto}, Todd C. and {Golimowski}, David A.},
	title = "{The Solar Neighborhood. VI. New Southern Nearby Stars Identified by Optical Spectroscopy}",
	journal = {\aj},
	year = 2002,
	month = apr,
	volume = {123},
	number = {4},
	pages = {2002-2009},
	doi = {10.1086/339315},
	archivePrefix = {arXiv},
	eprint = {astro-ph/0112496},
	primaryClass = {astro-ph},
	adsurl = {https://ui.adsabs.harvard.edu/abs/2002AJ....123.2002H}
}

@ARTICLE{OBrien2026.UV,
       author = {{O'Brien}, Mairi W. and {Tremblay}, Pier-Emmanuel and {G{\"a}nsicke}, Boris T. and {Hollands}, Mark A. and {Koester}, Detlev and {Sahu}, Snehalata and {B{\'e}dard}, Antoine and {Buchan}, Andrew M. and {Cunningham}, Tim and {Debes}, John H. and et al.},
        title = "{White dwarfs within 13 pc: insights from ultraviolet spectroscopy}",
      journal = {\mnras},
         year = 2026,
        month = oct,
       volume = {552},
       number = {1},
          eid = {stag1669},
        pages = {stag1669},
          doi = {10.1093/mnras/stag1669},
archivePrefix = {arXiv},
       eprint = {2607.06357},
 primaryClass = {astro-ph.SR},
       adsurl = {https://scixplorer.org/abs/2026MNRAS.552g1669O}
}

@ARTICLE{Mugrauer2005,
       author = {{Mugrauer}, M. and {Neuh{\"a}user}, R.},
        title = "{Gl86B: a white dwarf orbits an exoplanet host star}",
      journal = {\mnras},
         year = 2005,
        month = jul,
       volume = {361},
       number = {1},
        pages = {L15-L19},
          doi = {10.1111/j.1745-3933.2005.00055.x},
archivePrefix = {arXiv},
       eprint = {astro-ph/0506311},
 primaryClass = {astro-ph},
       adsurl = {https://ui.adsabs.harvard.edu/abs/2005MNRAS.361L..15M}
}

@ARTICLE{Farihi2013,
       author = {{Farihi}, J. and {Bond}, Howard E. and {Dufour}, P. and {Haghighipour}, N. and {Schaefer}, G.~H. and {Holberg}, J.~B. and {Barstow}, M.~A. and {Burleigh}, M.~R.},
        title = "{Orbital and evolutionary constraints on the planet hosting binary GJ 86 from the Hubble Space Telescope}",
      journal = {\mnras},
         year = 2013,
        month = mar,
       volume = {430},
       number = {1},
        pages = {652-660},
          doi = {10.1093/mnras/sts677},
archivePrefix = {arXiv},
       eprint = {1212.5601},
 primaryClass = {astro-ph.SR},
       adsurl = {https://ui.adsabs.harvard.edu/abs/2013MNRAS.430..652F}
}

@ARTICLE{Zeng2022,
       author = {{Zeng}, Yunlin and {Brandt}, Timothy D. and {Li}, Gongjie and {Dupuy}, Trent J. and {Li}, Yiting and {Brandt}, G. Mirek and {Farihi}, Jay and {Horner}, Jonathan and {Wittenmyer}, Robert A. and {Butler}, R. Paul. and {Tinney}, Christopher G. and {Carter}, Bradley D. and {Wright}, Duncan J. and {Jones}, Hugh R.~A. and {O'Toole}, Simon J.},
        title = "{The Gliese 86 Binary System: A Warm Jupiter Formed in a Disk Truncated at ≍2 au}",
      journal = {\aj},
         year = 2022,
        month = nov,
       volume = {164},
       number = {5},
          eid = {188},
        pages = {188},
          doi = {10.3847/1538-3881/ac8ff7},
archivePrefix = {arXiv},
       eprint = {2112.06394},
 primaryClass = {astro-ph.EP},
       adsurl = {https://ui.adsabs.harvard.edu/abs/2022AJ....164..188Z}
}

@ARTICLE{Alexander1969,
       author = {{Alexander}, J.~B. and {Lourens}, J. Van B.},
        title = "{The White Dwarf CoD -38$^{o}$10980}",
      journal = {Monthly Notes of the Astronomical Society of South Africa},
         year = 1969,
        month = jan,
       volume = {28},
        pages = {95},
       adsurl = {https://ui.adsabs.harvard.edu/abs/1969MNSSA..28...95A}
}

@ARTICLE{Luyten1949,
       author = {{Luyten}, Willem J.},
        title = "{An Atlas of Identification Charts of White Dwarfs.}",
      journal = {\apj},
         year = 1949,
        month = may,
       volume = {109},
        pages = {528},
          doi = {10.1086/145156},
       adsurl = {https://ui.adsabs.harvard.edu/abs/1949ApJ...109..528L}
}

@ARTICLE{Luyten1952,
       author = {{Luyten}, Willem J.},
        title = "{The Spectra and Luminosities of White Dwarfs.}",
      journal = {\apj},
         year = 1952,
        month = sep,
       volume = {116},
        pages = {283},
          doi = {10.1086/145612},
       adsurl = {https://ui.adsabs.harvard.edu/abs/1952ApJ...116..283L}
}
\bibliographystyle{mnras}



\end{document}